\documentclass[sigplan,screen,nonacm]{acmart}

\usepackage{amsmath,amsfonts,bm}
\usepackage{algorithmic}
\usepackage{algorithm}
\usepackage{array}
\usepackage{textcomp}
\usepackage{url}
\usepackage{verbatim}
\usepackage{graphicx}
\usepackage{xcolor}
\usepackage{tabularx} 
\usepackage{booktabs}
\usepackage{multirow}
\usepackage{pgfplots}
\pgfplotsset{compat=1.18}
\usepackage{tikz}
\usepackage{tcolorbox}
\usepackage{listings}
\usepackage{subcaption}

\usepgfplotslibrary{groupplots}
\usetikzlibrary{pgfplots.groupplots}
\usetikzlibrary{patterns, patterns.meta}
\usetikzlibrary{shapes.multipart, arrows.meta, positioning, fit}

\usetikzlibrary{arrows.meta, positioning}

\newcommand{\scheme}{\textsc{CacheMind}} 

\AtBeginDocument{%
  }

\setcctype{by}

\renewcommand\footnotetextcopyrightpermission[1]{}
\begin{document}

\title{\textsc{\scheme{}}:
From Miss Rates to Why — Natural-Language, Trace-Grounded Reasoning for Cache Replacement}

\renewcommand{\shorttitle}{\textsc{\scheme{}}:
Natural-Language, Trace-Grounded Reasoning for Cache Replacement}

\begin{abstract}
Cache replacement remains a challenging problem in CPU microarchitecture, often addressed using hand-crafted heuristics, limiting cache performance. 
Cache data analysis requires parsing millions of trace entries with manual filtering, making the process slow and non-interactive. 

To address this, we introduce \textsc{\scheme{}}, a conversational tool that uses Retrieval-Augmented Generation (RAG) and Large Language Models (LLMs) to enable semantic reasoning over cache traces. Architects can now ask natural language questions like, "\textit{Why is the memory access associated with PC X causing more evictions}?", and receive trace-grounded, human-readable answers linked to program semantics for the first time.

To evaluate \textsc{\scheme{}}, we present \textsc{\scheme{}Bench}, the first verified benchmark suite for LLM-based reasoning for the cache replacement problem. 
Using the \textsc{Sieve} retriever, \textsc{\scheme{}} achieves 66.67\% on 75 unseen trace-grounded questions and 84.80\% on 25 unseen policy-specific reasoning tasks; with \textsc{Ranger}, it achieves 89.33\% and 64.80\% on the same evaluations. Additionally, with \textsc{Ranger}, \textsc{\scheme{}} achieves 100\% accuracy on 4 out of 6 categories in the trace-grounded tier of \textsc{\scheme{}Bench}.  \textcolor{black}{Compared to LlamaIndex (10\% retrieval success), \textsc{Sieve} achieves 60\% and \textsc{Ranger} achieves 90\%, demonstrating that existing Retrieval-Augmented Generation (RAGs)  are insufficient for precise, trace-grounded microarchitectural reasoning.}

\textcolor{black}{ We provided four concrete actionable insights 
derived using \textsc{\scheme{}}, wherein bypassing use case improved cache hit rate by 7.66\% and speedup by 2.04\%, software fix use case gives speedup of 76\%, and Mockingjay replacement policy use case gives speedup of 0.7\%;
showing the utility of \textsc{\scheme{}} on non-trivial queries that require a natural-language interface. } 

\end{abstract}

\author{Kaushal Mhapsekar}
\orcid{0009-0006-3902-719X}
\affiliation{%
  \institution{North Carolina State University}
  \department{Electrical and Computer Engineering}
  \city{Raleigh}
  \state{North Carolina}
  \country{USA}
}
\email{kmhapse@ncsu.edu}

\author{Azam Ghanbari}
\orcid{0000-0002-4768-5663}
\affiliation{%
  \institution{North Carolina State University}
  \department{Electrical and Computer Engineering}
  \city{Raleigh}
  \state{North Carolina}
  \country{USA}
}
\email{aghanba2@ncsu.edu}

\author{Bita Aslrousta}
\orcid{0009-0009-5528-5342}
\affiliation{%
  \institution{North Carolina State University}
  \department{Electrical and Computer Engineering}
  \city{Raleigh}
  \state{North Carolina}
  \country{USA}
}
\email{baslrou@ncsu.edu}

\author{Samira Mirbagher-Ajorpaz}
\orcid{0009-0008-4997-5980}
\affiliation{%
  \institution{North Carolina State University}
  \department{Electrical and Computer Engineering}
  \city{Raleigh}
  \state{North Carolina}
  \country{USA}
}
\email{smirbag@ncsu.edu}

\renewcommand{\shortauthors}{Kaushal Mhapsekar, Azam Ghanbari, Bita Aslrousta, and Samira Mirbagher-Ajorpaz}




\maketitle

\section{Introduction}

Traditional cache replacement policies 
are predominantly crafted through manual design processes guided by empirical workload characterization, 
or brute-force exploration of the architectural parameter space ~\cite{dip, rrip, ship}. 
In contrast, recent state-of-the-art work learns replacement behavior directly from traces - \emph{Hawkeye} uses Belady-guided labels with an online classifier~\cite{hawkeye}, \emph{Glider} distills deep offline models into compact online policies~\cite{glider}, and \emph{PARROT} frames cache replacement as imitation learning with an offline solution~\cite{parrot}. Reinforcement-learning approaches such as \emph{RLR} and \emph{Stormbird} extend this line by incorporating bypass and access-type awareness~\cite{rlr,stormbird}. Complementary efforts study policy–prefetch interactions and cost effectiveness ~\cite{pacipv} and redesign at page granularity with translation-aware metadata (Genie Cache~\cite{kim2024genie}). Whereas, predictors like Hermes target long-latency loads that indirectly affect cache bottleneck~\cite{hermes}. Despite this progress, learned cache policies remain difficult to deploy in practice, often due to brittle contexts, opaque reasoning, and operationally heavy, limiting adoption in production systems.


Cache replacement analysis typically relies on fixed metrics (e.g., hit/miss rates) derived from the state-of-the-art processor simulators such as trace-driven ChampSim or cycle-accurate gem5,  and manual, offline trace inspection \cite{hawkeye, glider, rrip, ship} 
In practice, architects must sift through millions of events to answer simple “why” questions—\emph{which PC missed on which data, under which policy, and why}. This workflow is slow, non-interactive, and hard to trust when comparing policies across workloads.

Reasoning-focused evaluation relies on curated suites with unambiguous answers, such as \emph{GSM8K} (grade-school math), \emph{MATH} (competition math), \emph{DROP} (discrete reasoning over text), and \emph{ARC} (commonsense science)~\cite{gsm8k,math,drop,arc}. In architecture, \emph{ChampSim} and \emph{CRC-2} standardized trace-based evaluation for LLC policies~\cite{champsim}, but there is no community benchmark for \emph{trace-grounded reasoning}. Such a benchmark should: (i) ask questions about specific microarchitectural events (per-PC, per-address), (ii) tie answers to ground-truth traces, and (iii) score factual, comparative, and causal reasoning. This motivates our verified suite for cache-replacement analysis.



 We argue that the next generation of microarchitectural simulators must move beyond fixed metric outputs (e.g., hit/miss rates) to engines that can answer arbitrary, interactive questions about any event in the simulated execution. \textsc{\scheme{}} demonstrates this concept for cache replacement analysis, acting as a microarchitectural microscope that enables free-form, per-PC, per-data exploration.


We present \textsc{\scheme{}}, the first conversational, retrieval augmented system that turns raw ChampSim traces into a \emph{microarchitectural microscope}. Given a natural-language query, \textsc{\scheme{}} retrieves a narrow, verifiable slice of the trace (per-PC, per-address, per-event), augments it with policy/workload context and code snippets, and produces a grounded, human-readable answer. The goal is to support interactive, explainable reasoning about replacement behavior—not just aggregate reporting.

 Large language models (LLMs) are text predictors trained on vast corpora. In systems, they are increasingly used as \emph{tool users}: given a natural-language request, the model produces structured actions (e.g., code, queries, API calls) that read data, compute, and return an answer. Prior work (e.g., Toolformer, programmatic prompting frameworks) shows how to (i) ground an LLM with external tools and data, (ii) constrain its outputs to schemas or code, and (iii) verify the result against sources~\cite{toolformer,dspy}. For architecture tasks, this means translating questions like ``Which PC is causing most evictions under policy X?'' into precise database slices and simple analyses that can be checked.

RAG follows a ``\emph{search-then-write}'' recipe. Instead of asking a model to answer from memory, we first \emph{retrieve} a small set of relevant facts from an external store and then \emph{generate} the answer using only those facts. Retrieval can be keyword-based (sparse) or embedding-based (dense)~\cite{dpr}. Some architectures even retrieve during generation to keep long-range context fresh (e.g., RETRO)~\cite{retro}. Self-RAG variants let the model request or verify evidence mid-answer~\cite{selfrag}. In our setting, RAG boils millions of memory-trace entries down to a \emph{tiny}, verifiable window (e.g., all accesses to a specific PC and address under a given policy) that fits within the LLM’s context. This reduces hallucinations and makes answers checkable.

LLMs have limited context windows. To carry information across turns, systems maintain: (i) a \emph{sliding buffer} of recent messages; (ii) \emph{summaries} of older turns; and/or (iii) a \emph{vector store} (embedding index) of past facts that can be re-retrieved when similar questions arise. In effect, this is a \emph{conversation memory} layer. For trace analysis, buffers keep recent PCs, addresses, and policies handy; summaries help retain findings; and vector memory lets the system quickly rediscover relevant slices. External-memory LMs like RETRO perform a related trick during decoding by consulting an index rather than storing everything in the prompt~\cite{retro,memgpt}. 
We augmented the Generator LLM with conversation memory buffer, turning it into an assistive chat tool. This enables reasoning across multiple queries by retaining intermediate results, previous contexts, and trace-level insights.  

\begin{figure*}[!htbp]
  \centering
  \includegraphics[width=0.7\textwidth]{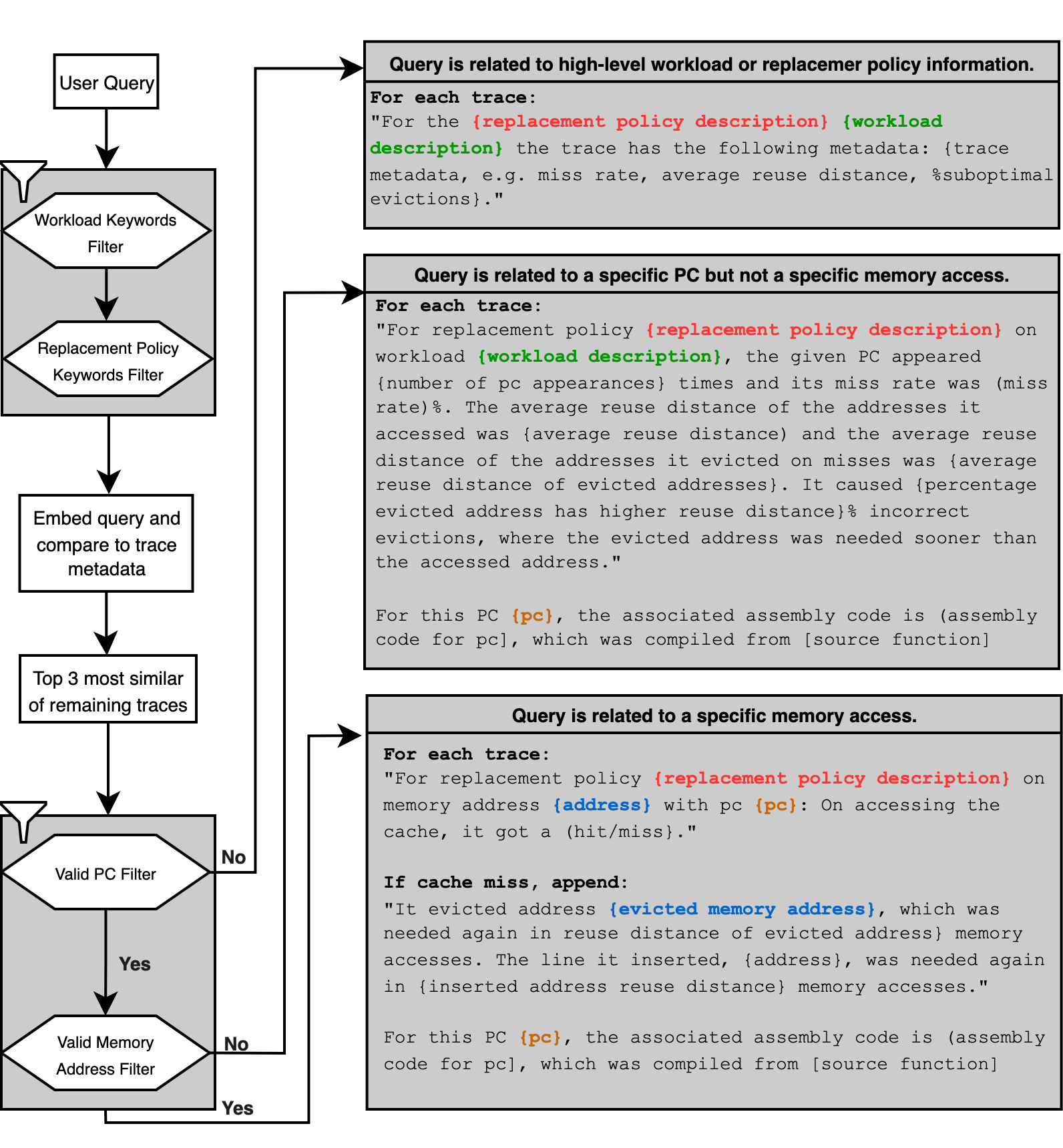}
   
  \caption{ 
  The method filters raw traces to a task-specific slice and returns the most informative evidence for the user’s query. Old ChampSim could tell you a miss; \textsc{\scheme{}} shows which PC missed on which data, under which policy, and why, for every event, acting as a microarchitectural \textit{microscope} that turns raw traces into per-PC, cross-policy answers.}
  \Description{Block diagram of the retrieval pipeline: query parsing, policy/workload selection, symbolic PC/address filtering, optional semantic ranking, aggregation of per-PC statistics and metadata, and assembly of a trace-grounded context for the LLM.}
  \label{fig:RAG_flowchart}
\end{figure*}

In \textsc{\scheme{}}, a user query is first parsed and routed to the \emph{retriever}, which searches an external store of trace-derived artifacts (e.g., per-PC slices, reuse statistics, and metadata). The retriever applies symbolic filters (policy, workload, PC, address) and, when applicable, semantic ranking to assemble a compact context bundle. This retrieved context conditions the \emph{generator} (an LLM), which produces a trace-grounded, context-aware response. Figure~\ref{fig:RAG_flowchart} presents a detailed overview of the architecture of the proposed retrieval system. 

We adopt \emph{retrieval-augmented generation (RAG)} to ensure answers are \emph{trace-grounded}. RAG serves two roles: (i) it fetches the exact slice(s) of simulator output from an external database needed to answer a query, and (ii) it supplies this retrieved context to the generator LLM so the response is informed by verifiable evidence rather than prior knowledge. In ablations where the generator is held fixed and only the retriever is toggled, enabling retrieval produces quantitatively grounded answers. All results reported in this paper are evaluated under the trace-grounded (RAG-enabled) setting.


To make progress measurable, we introduce \textsc{\scheme{} Bench}, a verified suite of 100 questions spanning factual retrieval, policy/workload comparison, arithmetic, adversarial “trick” checks, and high-level semantic analysis. Each item is trace-grounded with a single source of truth.

Full fine-tuning updates all model weights on domain data but is expensive. \emph{Parameter-efficient fine-tuning (PEFT)} (e.g., LoRA) inserts small adapters while freezing the base model, cutting trainable parameters by orders of magnitude~\cite{lora}. PEFT can inject domain style and terminology. However, if the new data are narrow, the model can become overconfident outside its niche (i.e., hallucinate more), so regularization and validation matter. 
We fine-tune GPT-4o-mini on domain-specific traces and also on prompts to improve precision for cache-specific queries while maintaining inference efficiency.  

Instead of updating weights, we can show 1–3 short examples in the prompt that demonstrate the task format or reasoning steps. This \emph{in-context learning} often boosts accuracy on structurally similar queries~\cite{gpt3}. \emph{Chain-of-thought} prompting asks the model to write intermediate steps, which can help verifiability on some tasks (with model- and task-specific gains)~\cite{cot}. 
\textsc{\scheme{}} performs one-shot and few-shot prompt engineering to improve LLM adaptability by passing one (one-shot) or three (few-shot) context-response example pairs to the Generator LLM.


Large Language Models (LLMs) exhibit \textit{emergent abilities}—capabilities that are not present in smaller models and do not result from explicit supervision, but appear once the model surpasses a critical threshold in scale~\cite{emergence} 
These include multi-step mathematical reasoning, logical deduction, and causal inference across benchmarks like GSM8K~\cite{gsm8k}, ARC~\cite{arc}, and DROP~\cite{drop}. 
Emergence allows models to generalize to tasks they were \textit{never explicitly trained to perform}.

Cache replacement policy reasoning exemplifies such a task. It requires tracing causality between temporally distant memory events, evaluating policy decisions across reuse distances, and linking low-level microarchitectural signals (e.g., PC, address, miss type) to high-level program semantics (e.g., function behavior, loop locality). This goes well beyond pattern classification or retrieval—it demands abstraction, explanation, and insight. No fixed heuristic or shallow model can answer arbitrary reasoning questions
across diverse workloads, replacement policies, and memory access patterns without supervised training on that. 

\textsc{\scheme{}} is the first to explicitly enhance multi-source, software-to-hardware causal reasoning, unleashing generative-AI to assist architects in understanding interactions that are currently invisible and enabling the first steps toward causal, phase-aware explanations of cache behavior. 
\textsc{\scheme{}} does not focus on simple facts that humans already know; but tasks that are hard for even experts to do. Modern caches exhibit complex cross-effects, prefetcher–replacement interactions (e.g., PACIPV ISCA’25) \cite{pacipv}, consistency/coherence behavior, inclusiveness, and page-size, that defy manual reasoning. 
In contrast, existing reasoning benchmarks such as QuArch \cite{quarch} ask static questions like “what translates a logical address into a physical?”, which reflect textbook knowledge rather than multi-layer causal analysis of caches, a complex dynamic system.  

In \textsc{\scheme{}}, the PC is a pointer to the line of code that must change in software or by the compiler, or to a common pattern that can be cheaply detected and handled in hardware e.g, a perceptron. PC-information assists with designing effective hash/signature functions of useful histories given the pool of addresses with minimal hardware overhead. This paper shows such actionable insights including software and hardware fixes such as how \textsc{\scheme{}} identifies bypass candidates and dead-on-arrival blocks and produces improved signatures to predict and bypass them,  identifying stable PCs with low reuse-distance variance to improve reuse-distance predictor training, guiding targeted software prefetch insertion by pinpointing dominant miss-causing instructions, and classifying hot and cold cache sets to support focused sampling and more effective replacement policy design.

We evaluate five variants of the proposed \textsc{\scheme{}} system, pairing its retrieval engine with different LLM backends: GPT-3.5-Turbo (legacy model), o3 (reasoning model), GPT-4o (flexible, general-purpose model), GPT-4o-mini (smaller, cheaper model), and a fine-tuned GPT-4o-mini(finetuned model). Performance is assessed across all eleven benchmark categories.  
Our results show \textsc{\scheme{}} framework successfully utilizes reasoning capability of LLMs for processor design queries.

%


\vspace{1em}
\noindent\textbf{Contribution.} This paper makes the following contributions: 


\begin{itemize}
  
  \item \textbf{\textsc{\scheme{}}:} a query-first, RAG-based framework for cache replacement analysis that answers per-event, per-PC questions with verifiable evidence. It introduced  a {\textit{Dual Retrieva}l} design, \emph{\textsc{Sieve}} (fast symbolic\allowbreak–semantic filtering) and \emph{\textsc{Ranger}} (LLM-based code retrieval) to balance precision and flexibility.
 
  \item \textbf{Benchmark:} \textsc{\scheme{}Bench}, 100 verified questions evaluating factual, comparative, arithmetic and semantic reasoning over traces.
 
  \item \textbf{Evaluation:} Using CRC-2 traces and multiple policies, \textsc{\scheme{}} achieves (90\%) accuracy on generic trace-based questions and (74.9\%) in reasoning tasks; \emph{\textsc{Ranger}} notably improves open-ended and arithmetic cases and provides fine-grain insights. \textsc{\scheme{}} also shows 9X better retrieval accuracy compared to LlamaIndex. 

  \item \textbf{Actionable Insights:} \textsc{\scheme{}} demonstrates utility by giving actionable insights such as examples provided:  1) bypass prediction, increasing hit rate by 7.66\% and IPC by 2.04\%, 2) enhanced Mockingjay RDP training, resulting in 0.7\% speedup and 3) software fixes giving 76\% speedup.

\end{itemize}

We argue that next-generation simulators should evolve from fixed reports to engines that can \emph{answer arbitrary, interactive questions} about any simulated event. \textsc{\scheme{}} demonstrates this direction for cache replacement. We plan to open-source  \textsc{\scheme{}Bench}, the retrieval pipeline, and artifacts to support reproducibility and follow-up work in the community.

\vspace{1em}
\noindent{\textbf{Paper organization.}}
Section 2 presents an overview of the retrieval-augmented system design and its two retrievers - \textsc{\scheme{}-Sieve} (symbolic–semantic filtering) and \textsc{\scheme{}-Ranger} (LLM-guided generation/execution)- along with the retrieval flow. 

Section 3 introduces \textsc{\scheme{}Bench}, detailing categories, scoring, and verification, and describes the external database/metadata that successfully ground queries (Figure~\ref{fig:system_prompt}, Figure~\ref{fig:trace_data}). Section 4 outlines methodology, including simulator configuration, workloads, and policies. 
Section 5 reports results: category-wise accuracy, the impact of retrieval quality, \textsc{Sieve} vs.\ \textsc{Ranger} comparison, LLM ablations (fine-tuning; one/few-shot prompting), model sensitivity, and qualitative observations. Section 6 discusses related work. Section 7 concludes and highlights future directions.


\section{Background and Related Works}

\paragraph{Heuristics-based policies.}
Classical cache replacement uses simple rules that approximate locality with tiny state and constant-time decisions. The most common is \emph{Least Recently Used (LRU)}, which evicts the line that has gone unused for the longest time; it works well when recently used data tends to be reused but breaks down on scans and weak temporal locality. \emph{Belady’s optimal} policy is an offline oracle that evicts the line whose \emph{next use} is farthest in the future; it defines the upper bound on hit rate but requires knowledge of the future and thus is not implementable in hardware~\cite{belady}. Practical online improvements include \emph{DIP} (Dynamic Insertion Policy), which uses \emph{set dueling} to mix insertion depths based on a few leader sets~\cite{dip}; \emph{RRIP/DRRIP}, which predicts a line’s re-reference interval with small saturating counters to resist scans and adapt across phases~\cite{rrip}; and \emph{SHiP}, which augments recency with simple program-context signatures to bias re-reference predictions~\cite{ship}. These families remain attractive due to their small area, simple control, and predictable timing.

\vspace{-0.5em}
\paragraph{Learning-based policies.}
Newer designs learn signals that better approximate Belady while staying deployable. \emph{Hawkeye} uses Belady-guided labels in sampled sets to classify cache-friendly lines and then applies that classifier online~\cite{hawkeye}. \emph{Glider} mines long-range structure with deep models offline and distills the behavior to a compact online policy~\cite{glider}. \emph{Mockingjay} extends learning-based replacement by predicting continuous reuse distance using PC-indexed temporal-difference learning, enabling eviction decisions that closely track Belady’s optimal ordering~\cite{mockingjay}. \emph{PARROT} treats replacement as \emph{imitation learning}, training an LSTM on traces to mimic an oracle and deploying a lightweight predictor~\cite{parrot}. These works show that information beyond pure recency (e.g., reuse-distance patterns and program context) helps at LLC scale.

\vspace{-0.5em}\paragraph{Machine-learning–inspired policies.}
Work adjacent to replacement has shaped model and feature choices for microarchitectural predictors. The \emph{perceptron branch predictor} demonstrated that linear models can exploit long histories efficiently in hardware~\cite{perceptron}. \emph{Multiperspective reuse prediction} aggregates features from program context, address bits, and control flow to guide bypass/promotion~\cite{mpppb}. Genetic search has also been applied to feature selection and weight tuning for compact classifiers~\cite{blbp}. In parallel, \emph{Reinforcement Learned Replacement (RLR)} shows how ML can craft competitive yet hardware-conscious replacement policies~\cite{rlr}, while \emph{Stormbird} recasts replacement as reinforcement learning, augmenting Belady-style guidance with bypass and access-type signals~\cite{stormbird}. These lines motivate compact models and robust featurization under tight latency/area budgets.

\vspace{-0.5em}\paragraph{LLM- and RAG-augmented tools.}Beyond replacement policies, a growing body of systems work augments domain tools with LLMs via retrieval-augmented generation. Healthcare assistants retrieve EHR/chart fragments to answer clinician queries with verifiable context \cite{ehr_ralm,medrag}; enterprise pipelines build graph-structured contexts prior to generation (GraphRAG) to improve faithfulness on large corpora \cite{graphrag}. MLOmics provides a curated, model-ready multi-omics repository with standardized schemas and programmatic access—well suited for RAG-style assistants that need verifiable, provenance-linked data \cite{mlomics}. These systems share our recipe: a domain index, a retriever that assembles task-specific evidence, and a generator constrained to ground its output in the retrieved context.

\section{\textsc{\scheme{}}}

\subsection{System Overview}




At the core of \textsc{\scheme{}} are two complementary retrievers and a generator:
(i) \emph{\textsc{Sieve}}, a symbolic–semantic filter that quickly isolates the right trace window from policy/workload/PC/address cues; and
(ii) \emph{\textsc{Ranger}}, an LLM-guided retriever that translates queries into executable code to fetch complex evidence when templates fall short.
A generator LLM then synthesizes a concise, trace-grounded explanation.


\paragraph{Retrieval} \textsc{\scheme{}} employs two complementary retrieval strategies: 
\begin{enumerate}
    \item \textsc{\scheme{}-Sieve}, a filter-based retriever that combines symbolic and semantic filtering to efficiently extract relevant trace entries for LLM interpretation (Figure~\ref{fig:RAG_flowchart}), and 
    \item \textsc{\scheme{}-Ranger}, an LLM-based retriever that leverages large language models to dynamically generate and execute retrieval queries against the external database (Figure~\ref{fig:system_prompt}).
\end{enumerate}

\subsection{\textsc{\scheme{}-Sieve} (Symbolic-Indexed Entries for Verifiable Extraction)}
The \textsc{\scheme{}-Sieve} retriever uses a multi-stage process tailored for ChampSim traces. 

Figure ~\ref{fig:RAG_flowchart} depicts \textsc{\scheme{}-Sieve} architecture and the retrieval flow. It illustrates how \textsc{\scheme{}}, filters raw traces to a task-specific slice and returns the most informative evidence for the user’s query. It contains the block diagram of the retrieval pipeline: query parsing, policy/workload selection, symbolic PC/address filtering, optional semantic ranking, aggregation of per-PC statistics and metadata, and assembly of a trace-grounded context for the LLM.
 

\subsubsection{Trace-Level Filtering Module.}
The first stage applies a sentence embedder to extract workload and replacement policy names from the user query. These are ranked by similarity and matched against the database keys to narrow the search space. For example:
\begin{itemize}
    \item Input query: ``What is the miss rate for PC 0x4037ba on the mcf workload with PARROT replacement policy?''
    \item Translated query: filter by \texttt{policy=PARROT}, \texttt{workload\allowbreak=mcf}
    \item Output: corresponding trace slice from the key \texttt{mcf\allowbreak\_evictions\allowbreak\_parrot}
\end{itemize}

\subsubsection{PC and Address Filtering Module.}
The second stage applies symbolic filters on program counters (PCs) and memory addresses if present in the query. For example:

\begin{itemize}
    \item Query: ``Does the memory access with PC 0x401e31 and address 0x35e798a637f result in a cache hit or miss for the lbm workload under PARROT?''
    \item Filters applied: \texttt{program\_counter == 0x401e31}, \texttt{memory\allowbreak\_address == 0x35e798a637f}
    \item Result: isolates a compact trace slice from millions of entries in \verb|lbm_evictions_parrot|.
\end{itemize}
Each database entry stores PC, address, workload, policy, eviction type, and metadata such as recency, reuse distance, and cache state.

\subsubsection{Cache Statistical Expert.}
For PCs present in the retrieved slice, helper functions compute statistics including miss rate, access and eviction reuse distances, and percentage of bad evictions.

\subsubsection{Response Generation LLM Module.}
The final response is assembled by combining workload information, policy description, PC-level context (function name, assembly, cache statistics), and trace metadata. This structured response grounds the subsequent LLM interpretation. An example of the prompt is shown in Figure ~\ref{fig:trace_data}

\begin{figure}[!htbp]
\centering
\setlength{\fboxsep}{6pt}
\setlength{\fboxrule}{0.5pt}
\fbox{%
  \begin{minipage}{0.95\linewidth}
  \ttfamily\footnotesize\raggedright

  \textbf{Cache Access Trace} \\[0.25em]
  \begin{itemize}
    \item \textbf{PC:} \texttt{0x405832}
    \item \textbf{Address:} \texttt{0x2a9e6a48d9d}
    \item \textbf{Set ID:} \texttt{0b10110011101}
    \item \textbf{Evict:} \texttt{true}
  \end{itemize}
  \rule{\linewidth}{0.4pt}

  \textbf{Cache Lines} \\[-0.25em]
  \begin{itemize}
    \item \{\texttt{"0x2c919839d9d"}, \texttt{"0x405832"}\}
    \item \{\texttt{"0x2c91983e59d"}, \texttt{"0x405832"}\}
    \item \dots
  \end{itemize}

  \textbf{Access History} \\[-0.25em]
  \begin{itemize}
    \item \{\texttt{"0x3528e1a7d9d"}, \texttt{"0x409228"}\}
    \item \{\texttt{"0x286af5fd59d"}, \texttt{"0x409270"}\}
    \item \dots
  \end{itemize}

  \textbf{Cache Line Scores} \\[-0.25em]
  \begin{itemize}
    \item \{\texttt{3062739738013}, \texttt{180}\},
          \{\texttt{3062739756445}, \texttt{181}\},
          \{\texttt{3707401110941}, \texttt{195}\}
  \end{itemize}
  \rule{\linewidth}{0.4pt}

  \textbf{Assembly (mainSimpleSort)} \\[-0.25em]
  \begin{tabbing}
    405821:\quad \= 84 c0 \quad \= test \quad \= \%al,\%al \\
    405832:\quad \= 0f 85 \> jne \> 4032d7 <mainSimpleSort+0xbd> \\
    405839:\quad \= eb 01 \quad \> jmp \> 40336d <mainSimpleSort+0x153> \\
    40583b:\quad \= 90 \quad \> nop \> \\
    40583f:\quad \= 8b 45 ec \> mov \> -0x14(\%rbp),\%eax
  \end{tabbing}

  \end{minipage}%
}
\vspace{-1em}\caption{Example trace excerpt retrieved by \textsc{\scheme{}} }
\Description{Monospaced listing of a cache-access record with PC, address, set ID, recent cache lines and access history, an eviction flag, simple per-line scores, and a short disassembly around the PC. }
\label{fig:trace_data}
\end{figure}


\subsection{\textsc{\scheme{}-Ranger} (Retrieval via Agentic Neural Generation and Execution Runtime)}

While \textsc{\scheme{}-Sieve} achieves high precision for structured \allowbreak queries, it is less effective for open-ended or dynamic queries. To address this limitation, \textsc{\scheme{}} introduces \textsc{\scheme{}-Ranger}, an LLM-based retriever. The Retrieval LLM is provided with:
\begin{itemize}
    \item The retrieval objective (e.g., generate structured code to query the database),
    \item A detailed schema of the external database, and
    \item Task-specific instructions and formatting rules.
\end{itemize}
The model uses the system prompt, shown in Figure~\ref{fig:system_prompt}, to generate executable Python code that extracts relevant entries from the external trace database. This design enables handling of queries that deviate from predefined templates and leverages the LLM’s ability to map natural language requests into structured data retrieval. We have used OpenAI's GPT-4o model for this purpose.

\begin{figure}[!t]
\centering
\setlength{\fboxsep}{6pt}
\setlength{\fboxrule}{0.5pt}
\fbox{%
  \begin{minipage}{0.98\linewidth}
  \ttfamily\scriptsize\raggedright
  \textbf{SYSTEM PROMPT} \\[0.25em]
  You are a Python code-writing assistant for analyzing cache memory trace data.
  Your task is to generate Python code that extracts \textbf{string-formatted answers}
  from a dictionary named \texttt{loaded\_data}. \\[0.25em]
  \rule{\linewidth}{0.4pt}

  \textbf{Data Structure Overview} \\[-0.25em]
  \begin{itemize}
    \setlength{\itemsep}{1pt}\setlength{\parsep}{0pt}\setlength{\parskip}{0pt}
    \item \texttt{loaded\_data}: a dictionary with keys like
          \texttt{lbm\_\allowbreak evictions\_\allowbreak lru}.
    \item Values: \texttt{"data\_frame"} (pandas DataFrame),
          \texttt{"metadata"} (string), \texttt{"description"} (string).
    \item Workloads: \texttt{astar}, \texttt{lbm}, \texttt{mcf}.
    \item Policies: \texttt{belady}, \texttt{lru}, \texttt{mlp}, \texttt{parrot}.
  \end{itemize}

  \textbf{Dataframe Structure (\texttt{data\_frame})} \\[-0.25em]
  Columns include (non-exhaustive):
  \begin{itemize}
    \setlength{\itemsep}{1pt}\setlength{\parsep}{0pt}\setlength{\parskip}{0pt}
    \item \texttt{program\_counter}, \texttt{memory\_address}, \texttt{cache\_set\_id},
          \texttt{evict} (\texttt{Cache Hit}/\texttt{Cache Miss}), \texttt{miss\_type}
    \item reuse/recency features, \texttt{function\_code},
          \texttt{assembly\_code}
    \item \texttt{current\_cache\_lines}, \texttt{recent\_access\_history},
          \texttt{cache\_line\_eviction\_scores}
  \end{itemize}
  \rule{\linewidth}{0.4pt}

  \textbf{Metadata (\texttt{metadata})} \\[-0.25em]
  \begin{itemize}
    \setlength{\itemsep}{1pt}\setlength{\parsep}{0pt}\setlength{\parskip}{0pt}
    \item A single string summarizing trace stats (accesses, misses, evictions,
          miss rate, correlations, etc.).
    \item Access via \texttt{loaded\_data[trace\_id]["metadata"]}.
    \item Extract numbers with simple matching or regex, e.g.\ \texttt{re.search(r"total misses", metadata)}.
    \item Example:
    \begin{quote}\ttfamily
      Cache Performance Summary: 140704 total accesses, 133542 total misses, 94.91\% miss rate,\\
      100.00\% capacity misses, 0.00\% conflict misses, 133478 total evictions,\\
      87085 (65.24\%) wrong evictions where evicted line has lower reuse distance.\\
      The correlation between accessed address recency and cache misses is 0.18.
    \end{quote}
  \end{itemize}
  \rule{\linewidth}{0.4pt}

  \textbf{Task Instructions} \\[-0.25em]
  \begin{itemize}
    \setlength{\itemsep}{1pt}\setlength{\parsep}{0pt}\setlength{\parskip}{0pt}
    \item First check matching \texttt{workload}/\texttt{policy}; then check \texttt{PC}/\texttt{address};
          finally fall back to \texttt{metadata}.
    \item Return a single result string with hit/miss, reuse/recency, relevant metadata summary,
          and assembly context.
    \item If nothing is found, return a clear message.
  \end{itemize}

  \textbf{Output Rules} \\[-0.25em]
  \begin{itemize}
    \setlength{\itemsep}{1pt}\setlength{\parsep}{0pt}\setlength{\parskip}{0pt}
    \item Must set \texttt{result = "..."} (a Python string).
    \item No markdown, explanations, \texttt{print}, or comments.
  \end{itemize}

  \textbf{Valid Examples} \\
  \quad \texttt{result = f"The miss rate for PC 0x401e31 is 44.69\%."} \\[0.25em]
  \textbf{Invalid Examples} \\
  \quad \texttt{return df["miss\_rate"]}, \texttt{print(result)}, \texttt{result = df}
  \end{minipage}%
}
\vspace{-1em}\caption{ Formatted system prompt summarizing the data container, dataframe schema, metadata string, task flow, and strict output rules for code generation with \textsc{\scheme{}-Ranger}.  
The ability to compare, find the root cause of performance differences among designs, and to judge replacement policies accurately, is the essential first step toward automating cache policy debugging and design assistant using AI.  }
\Description{Formatted system prompt summarizing the data container, dataframe schema, metadata string, task flow, and strict output rules for code generation. }
\label{fig:system_prompt}
\end{figure}

\section{\textsc{\scheme{}Bench}: Novel Verified Benchmark Suite for AI Evaluation}

\newcolumntype{C}[1]{>{\centering\arraybackslash}p{#1}}
\providecommand{\catline}{\cline{2-5}} 

\begin{table*}[!htbp]
  \label{tab:benchmark_questions}
  \centering
  \small
  \renewcommand{\arraystretch}{1.1}
  \begin{tabularx}{\textwidth}{|C{2.25cm}|p{2.75cm}|p{4cm}|c|X|}
    \hline
    \textbf{Benchmark} & \textbf{Category} & \textbf{Description} & \textbf{\#} & \textbf{Representative Example} \\
    \hline

    \multirow[t]{6}{2.8cm}{\emph{Trace-Grounded Questions (75)}}%
      & Cache Hit/Miss
      & Determine if a specific access results in a hit or miss
      & 30
      & ``Does PC 0x401dc9 and address 0x47ea85d37f result in a cache hit in \texttt{lbm} under \texttt{PARROT}?'' \\
    \catline
    & Miss Rate
      & Compute miss rate of a specific PC or workload
      & 10
      & ``What is the miss rate for PC 0x4037ba in \texttt{mcf} with \texttt{PARROT}?'' \\
    \catline
    & Policy Comparison
      & Compare miss behavior across policies
      & 15
      & ``Which policy has the lowest miss rate for PC 0x409270 in \texttt{astar}?'' \\
    \catline
    & Count
      & Count frequency of events (e.g., accesses, evictions)
      & 5
      & ``How many times did PC 0x405832 appear in \texttt{astar} under \texttt{LRU}?'' \\
    \catline
    & Arithmetic
      & Perform arithmetic over trace statistics
      & 10
      & ``What is the average \emph{evicted reuse distance} of PC 0x40170a for the \texttt{lbm} workload with \texttt{MLP}?'' \\
    \catline
    & Trick Question
      & Catch inconsistencies or invalid assumptions
      & 5
      & ``Does PC 0x4037aa in \texttt{lbm} access address 0x1b73be82e3f?'' (Answer: \emph{TRICK}) \\
    \hline

    \multirow[t]{5}{2.8cm}{\emph{Architectural Reasoning \\ and Analysis \allowbreak (25)}}%
      & Microarchitecture Concepts
      & Answer general cache questions with logic
      & 5
      & ``How does increasing cache size affect miss rate? Compare increasing \#sets vs.\ \#ways.'' \\
    \catline
    & Code Generation
      & Generate code to analyze specific trace conditions
      & 5
      & ``Write code to compute hits for PC 0x4037ba and address 0xa3a0df3d9d in \texttt{mcf} under \texttt{LRU}.'' \\
    \catline
    & Replacement Policy
      & Analyze behavior of policies across contexts
      & 5
      & ``Why does Belady outperform \texttt{LRU} on PC 0x409270 in \texttt{astar}?'' \\
    \catline
    & Workload Analysis
      & Reason about entire workload characteristics
      & 5
      & ``Which workload has the highest cache miss rate under \texttt{MLP}?'' \\
    \catline
    & Semantic Analysis
      & Connect PCs to high-level program behavior
      & 5
      & ``Why does PC 0x4037ba have a high hit rate? Examine the assembly context and analyze.'' \\
    \hline
  \end{tabularx}
  \caption{  \textsc{\scheme{}Bench} categories and representative queries
  with counts and example questions for each.}
  \Description{Categories in \scheme{}Bench with counts and example questions for each. }
  \label{tab:benchmark_questions}
\end{table*}

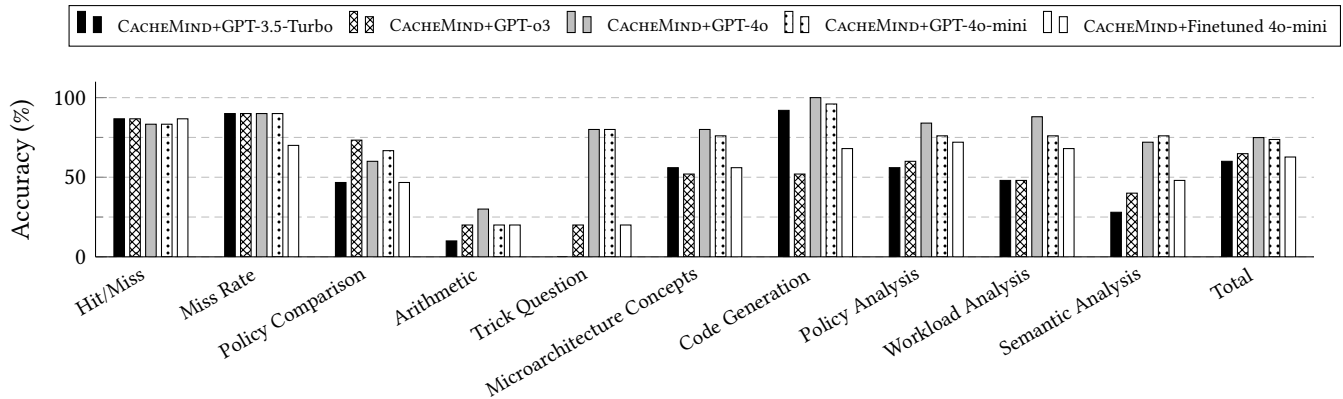
\begin{figure*}[!t]
\centering
\begin{tikzpicture}
\begin{axis}[
    ybar,
    bar width=4pt,
    width=\textwidth,
    height=0.22\textwidth,
    ymin=0,
    ymax=110,
    ylabel={Accuracy (\%)},
    symbolic x coords={
        Hit/Miss, Miss Rate, Policy Comparison, Arithmetic, Trick Question,
        Microarchitecture Concepts, Code Generation, Policy Analysis, Workload Analysis, Semantic Analysis, Total
    },
    xtick=data,
    xtick style={draw=none},
    x tick label style={rotate=30, anchor=east, font=\footnotesize},
    y tick label style={font=\small},
    axis x line*=bottom,
    axis y line*=left,
    enlarge x limits=0.05,
    ymajorgrids,
    yminorgrids,
    minor y tick num=1,
    y grid style={densely dashed, gray!50},
    minor y grid style={densely dashed, gray!50},
    legend style={
        at={(0.5,1.2)}, anchor=south,
        legend columns=5,
        column sep=4pt,
        font=\scriptsize
    }
]

\addplot+[fill=black, draw=black] coordinates {
    (Hit/Miss,86.7) (Miss Rate,90) (Policy Comparison,46.7)
    (Arithmetic,10) (Trick Question,0) 
    (Microarchitecture Concepts,56) (Code Generation,92) (Policy Analysis,56)
    (Workload Analysis,48) (Semantic Analysis,28) (Total,60.0)
};

\addplot+[pattern=crosshatch, pattern color=black, draw=black] coordinates {
    (Hit/Miss,86.7) (Miss Rate,90) (Policy Comparison,73.3)
    (Arithmetic,20) (Trick Question,20)
    (Microarchitecture Concepts,52) (Code Generation,52) (Policy Analysis,60)
    (Workload Analysis,48) (Semantic Analysis,40) (Total,64.8)
};

\addplot+[fill=lightgray, draw=black] coordinates {
    (Hit/Miss,83.3) (Miss Rate,90) (Policy Comparison,60)
    (Arithmetic,30) (Trick Question,80) 
    (Microarchitecture Concepts,80) (Code Generation,100) (Policy Analysis,84)
    (Workload Analysis,88) (Semantic Analysis,72) (Total,74.9)
};

\addplot+[pattern=dots,pattern color=black, draw=black] coordinates {
    (Hit/Miss,83.3) (Miss Rate,90) (Policy Comparison,66.7)
    (Arithmetic,20) (Trick Question,80) 
    (Microarchitecture Concepts,76) (Code Generation,96) (Policy Analysis,76)
    (Workload Analysis,76) (Semantic Analysis,76) (Total,73.7)
};

\addplot+[
    pattern=crosshatch,
    pattern color=black,
    fill=white,
    draw=black
] coordinates {
    (Hit/Miss,86.7) (Miss Rate,70) (Policy Comparison,46.7)
    (Arithmetic,20) (Trick Question,20) 
    (Microarchitecture Concepts,56) (Code Generation,68) (Policy Analysis,72)
    (Workload Analysis,68) (Semantic Analysis,48) (Total,62.7)
};

\legend{
    \scheme{}+GPT-3.5-Turbo,
    \scheme{}+GPT-o3,
    \scheme{}+GPT-4o,
    \scheme{}+GPT-4o-mini,
    \scheme{}+Finetuned 4o-mini
}

\end{axis}
\end{tikzpicture}
\caption{Accuracy of \textsc{\scheme{}} with different LLM backends across \textsc{\scheme{}Bench} categories.}
\Description{}
\label{fig:main}
\end{figure*}
\textsc{\scheme{}Bench} comprises 100 verified, trace-grounded questions in two tiers designed to isolate retrieval accuracy from higher-level reasoning:
\begin{itemize}
  \item \textbf{Trace-Grounded (TG-QA) Questions (75).} Assess precise retrieval and basic logic, e.g., hit/miss classification, per-PC miss rates, cross-policy comparisons, counting, and simple arithmetic. Example: ``Does the access with PC \verb|0x401dc9| and address \verb|0x47ea85d37f| result in a cache hit under \texttt{LRU} on \texttt{lbm}?'' 
  \item \textbf{Architectural Reasoning and Analysis (ARA) Questions (25).} Evaluate analytical and reasoning capability when provided trace context (e.g., policy/workload descriptions, per-PC slices, statistics). Example: ``Why does PC \verb|0x401e31| perform worse under \texttt{PARROT} than \texttt{Belady} on \texttt{lbm}?'' 
\end{itemize}

\subsection{Trace-Grounded tier.}
All questions are grounded in simulator outputs and verified against trace logs. This tier targets fine-grained retrieval fidelity at PC/address granularity and includes:
\begin{itemize}
  \item \textbf{Per-access verification:} Hit/miss classification for specific \{\texttt{PC}, \texttt{address}, \texttt{policy}, \texttt{workload}\} tuples.
  \item \textbf{Per-PC miss rate analysis:} Miss-rate breakdowns per PC, often across multiple policies.
  \item \textbf{Cross-policy ranking:} Ordering policies by hit/miss behavior per PC or workload.
  \item \textbf{Counting:} Frequency of events (e.g., accesses, evictions) under specified filters.
  \item \textbf{Arithmetic:} Simple computations over retrieved fields (e.g., mean evicted reuse distance).
  \item \textbf{Trick questions:} Premise checks that should be rejected (e.g., mismatched PC/workload), probing hallucination and uncertainty handling.
\end{itemize}
Scoring is binary (0/1) using exact-match verification against the trace.

\subsection{Architectural Reasoning and Analysis (ARA) tier.}
This subset targets trace-informed analysis and explanation, evaluating how well the system synthesizes retrieved and is categorized into:
\begin{itemize}
  \item \textbf{Microarchitecture concepts:} Retrieval-light questions probing general understanding of cache behavior.
  \item \textbf{Code generation:} Requests for brief analysis code operating on the specified trace slice.
  \item \textbf{Policy analysis:} Explanations that link policy mechanics (e.g., recency vs.\ reuse distance) to observed PC-level effects.
  \item \textbf{Workload analysis:} Comparisons across workloads given their descriptions and retrieved metadata.
  \item \textbf{Semantic analysis:} Linking trace statistics to disassembly/context at a PC to explain higher-level behavior.
\end{itemize}

Answers are rubric-graded on a 0–5 scale for correctness, use of evidence, and clarity.

Table~\ref{tab:benchmark_questions} summarizes the 11 categories across both tiers and their sizes in the validated dataset.

\subsection{External Database and Metadata}

\paragraph{Data organization.}
The external store is a Python dictionary, \texttt{loaded\_data}, whose keys are trace identifiers of the form
\texttt{\textless workload\textgreater\_\allowbreak evictions\_\allowbreak \textless policy\textgreater}
(e.g., \texttt{lbm\_\allowbreak evictions\_\allowbreak lru}). Each key maps to a dictionary with three fields:
\begin{itemize}
  \item \texttt{data\_frame}: a \emph{pandas DataFrame} containing per-access records (not a list of dicts).
  \item \texttt{metadata}: a single descriptive string summarizing whole-trace statistics (e.g., totals, rates, correlations).
  \item \texttt{description}: a short human-readable summary of the workload and policy.
\end{itemize}


\paragraph{DataFrame schema.}
Each \texttt{data\_frame} follows a uniform schema; below we list each column and its purpose:
\begin{itemize}
  \item \texttt{program\_counter} - Instruction identity (e.g., \texttt{0x401d9b}).
  \item \texttt{memory\_address} - Accessed memory location (e.g., \texttt{0x35e798a637f}).
  \item \texttt{cache\_set\_id} - Target cache set.
  \item \texttt{evict} - Access outcome (\emph{Cache Hit}/\emph{Cache Miss}).
  \item \texttt{miss\_type} - Miss taxonomy (e.g., \emph{Capacity}, \emph{Conflict}).
  \item \texttt{evicted\_address} - Line evicted by this access (if any).
  \item \texttt{accessed\_address\_recency} - Textual recency descriptor.
  \item \texttt{accessed\_address\_reuse\_distance} - Reuse distance for the accessed line.
  \item \texttt{evicted\_address\_reuse\_distance} - Reuse distance for the evicted line.
  \item \texttt{function\_name} - Source-level function name mapped from PC.
  \item \texttt{function\_code} - Short source snippet around the PC.
  \item \texttt{assembly\_code} - Disassembly around the PC.
  \item \texttt{current\_cache\_lines} - Snapshot of (PC, address) pairs resident in the set at access time.
  \item \texttt{recent\_access\_history} - Recent (PC, address) tuples for context.
  \item \texttt{cache\_line\_eviction\_scores} - Per-line scores used by the policy to decide evictions.
  \item \texttt{current\_cache\_line\_addresses} - Addresses resident in the set at access time.
  \item \texttt{evicted\_address\_reuse\_distance\_numeric} - Reuse distance for the evicted line.
  \item \texttt{accessed\_address\_reuse\_distance\_numeric} - Reuse distance for the accessed line.
  \item \texttt{accessed\_address\_recency\_numeric} -Access recency (number of intervening accesses).
  \item \texttt{is\_miss} - Indicator for miss/hit (1 = miss, 0 = hit).
\end{itemize}

\paragraph{Trace-level metadata.}
The \texttt{metadata} field is a single free-form string summarizing the entire trace (e.g., total accesses/misses/evictions,
miss rate, miss breakdowns, wrong-eviction ratios, correlations). It is accessed as
\texttt{loaded\_data[\allowbreak trace\_id][\allowbreak "metadata"]} and parsed with string matching or regular expressions when numeric
values are needed. A representative snippet:
\emph{"Cache Performance Summary: 140{,}704 total accesses, 133{,}542 total misses, 94.91\% miss rate; 100.00\% capacity misses,
0.00\% conflict misses; 133{,}478 total evictions; 87{,}085 (65.24\%) wrong evictions where the evicted line has lower reuse distance.
Correlation between accessed-address recency and cache misses: 0.18."}


\section{Methodology}

\paragraph{Simulator.} 
For LLC trace generation, we use ChampSim~\cite{champsim}, a trace-based simulator widely used in prior work and by CRC-2. Building on the PARROT infrastructure~\cite{parrot,google_research_cache_replacement}, we replay LLC accesses and emit eviction-annotated traces under Belady's optimal, LRU, and PARROT. To broaden the comparison within the same framework, we implemented and integrated an MLP-based replacement policy ourselves. The resulting traces are normalized and stored as the external database used in our experiments.  The processor and memory hierarchy used in all experiments is summarized in Table~\ref{table:arch_config}.


We also extended \textsc{\scheme{}} to the cycle-accurate gem5 simulator to support a broader range of software interventions as use cases for \scheme{}, as well as testing other workloads and trace sources, enabling access to richer microarchitectural features such as complex instructions, access types, and cache-level information. This gem5-based version of \textsc{\scheme{}} is used to evaluate software-level interventions, including software prefetching for high missed PCs, as reported in Section~\ref{sec:actionable}.



\begin{table}[!htbp]
\footnotesize
\centering

\begin{tabularx}{\columnwidth}{@{}lX@{}}
\toprule
\textbf{Component} & \textbf{Configuration} \\
\midrule
Processor & 1 core; 4 GHz; 6-wide fetch/decode/execute; 4-wide retire; 352-entry ROB; 128-entry LQ; 72-entry SQ; bimodal branch predictor \\
L1 I-Cache & 32 KB, 64 sets, 8 ways; 4-cycle latency; 8-entry MSHR; LRU \\
L1 D-Cache & 32 KB, 64 sets, 8 ways; 4-cycle latency; 16-entry MSHR; LRU \\
L2 Cache & 512 KB, 1024 sets, 8 ways; 12-cycle latency; 32-entry MSHR; LRU \\
LLC & 2 MB, 2048 sets, 16 ways; 26-cycle latency; 64-entry MSHR; Replacement Policies - Belady's optimal, LRU, PARROT, Multi-layer Perceptron \\
DRAM & 4 GB; DDR4-3200MT/s; 1 channel; 1 rank/channel; 8 banks/rank; 512 MB/bank; 8-bit channel; tRP = tRCD = tCAS = 12.5 ns \\
\bottomrule
\end{tabularx}
\caption{Processor and Memory Configuration.} 
\label{table:arch_config}
\end{table}
\vspace{-0.5em}
\paragraph{Workloads.} 
We evaluate the \texttt{astar}, \texttt{lbm} and \texttt{mcf} workloads derived from SPEC CPU 2006 benchmarks used in CRC-2, producing detailed memory traces. Each trace records program counters (PCs), memory addresses, cache hit/miss outcomes, and eviction decisions. Simulations are executed for each workload with 0 warm-up instructions and 1 billion simulation instructions. We set the warm-up phase to zero and simulate the entire workload traces, as our goal is to analyze cache-access events (hits/misses, evictions, reuse/recency) rather than steady-state performance statistics.
\vspace{-0.5em}
\paragraph{Replacement Policies.}
Cache behavior is measured under both baseline and learned policies. Baseline policies include Least Recently Used (LRU) and Belady’s optimal, while learned policies include PARROT~\cite{parrot} and a Multi-Layer Perceptron (MLP)~\cite{mpppb}. The OpenAI Gym framework ~\cite{google_research_cache_replacement} is used to simulate these policies and generate metadata. We extend the framework to support MLP in addition to the existing LRU, Belady, and PARROT implementations, enabling direct comparison of heuristic, oracle, and neural approaches in a unified environment.
\vspace{-0.5em}
\paragraph{Traces and Metadata.} 
To enrich trace semantics, we augment ChampSim outputs with source-level metadata. Each PC is linked to its corresponding assembly and source code, while additional cache parameters (e.g., recency, reuse distance, and PC--address correlations) are extracted to form feature sets for machine learning. The resulting database comprises 4.52 GB of data, covering three workloads and four replacement policies, yielding twelve dataframes in total, which is approximately 375 MB per dataframe (1 workload and 1 policy). An extended database with potentially 8-10 replacement policies and 30 workloads could be roughly 100 GB large to be analyzed.  

\section{Results}

\subsection{Evaluation by Query Category \label{sec_6_1}}

Figure~\ref{fig:main} shows the  breakdown results by the category of queries in \textsc{\scheme{}Bench}. 




\vspace{-0.5em}
\paragraph{Cache Hit/Miss.}
All models are strong when the retrieval context is high. With 30 queries, GPT-4o and GPT-4o-mini reach 83.3\%, while GPT-3.5, o3, and finetuned 4o-mini each score 86.7\%. These are direct lookups once the PC+address+ policy+workload tuple is isolated.

\vspace{-0.5em}
\paragraph{Miss Rate.}
On 10 queries, GPT-3.5/o3/GPT-4o/GPT-4o-mini achieve 90.0\%; the finetuned 4o-mini yields 80.0\%. Computing a per-PC or per-workload miss rate remains easy when the relevant slice is retrieved.

\vspace{-0.5em}
\paragraph{Policy Comparison.}
When asked to compare policies for a given PC or workload (15 queries), performance diverges: o3 leads at 73.3\%, GPT-4o-mini at 66.7\%, GPT-4o at 60.0\%, while GPT-3.5 and finetuned 4o-mini are at 46.7\%. These require selecting and ranking across multiple retrieved statistics rather than returning a single value.

\vspace{-0.5em}
\paragraph{Count.}
All models score 0/5. Pure counting over low-context windows is brittle: a single missed filter or failure to iterate the entire slice yields an incorrect result.

\vspace{-0.5em}
\paragraph{Arithmetic.}
Numerical aggregation beyond simple rates remains challenging with low context (10 queries): GPT-4o reaches 30.0\%, GPT-3.5 10.0\%, o3/GPT-4o-mini/finetuned 4o-mini each 20.0\%. These typically require multiple filtered values and a secondary computation (e.g., averaging reuse distances).

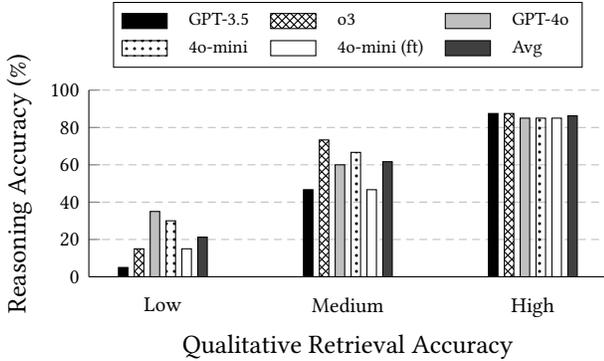
\begin{figure}[!htbp]
  \centering
  \begin{tikzpicture}
    \begin{axis}[
      ybar,
      bar width=3.6pt,
      width=\linewidth,
      height=0.48\linewidth,
      enlarge x limits=0.20,
      ymin=0, ymax=100,
      ylabel={Reasoning Accuracy (\%)},
      xlabel={Qualitative Retrieval Accuracy},
      symbolic x coords={Low, Medium, High},
      xtick=data,
      xtick style={draw=none},
      x tick label style={font=\footnotesize},
      y tick label style={font=\footnotesize},
      axis x line*=bottom,
      axis y line*=left,
      ymajorgrids,
      y grid style={densely dashed, gray!50},
      legend style={
        at={(0.5,1.12)}, anchor=south,
        legend columns=3,
        column sep=6pt,
        font=\scriptsize
      },
      legend cell align={left},
    ]

    \addplot+[fill=black, draw=black, bar shift=-15pt, area legend]
      coordinates {(Low,5.0) (Medium,46.67) (High,87.5)};

    \addplot+[pattern=crosshatch, pattern color=black, draw=black, bar shift=-9pt, area legend]
      coordinates {(Low,15.0) (Medium,73.33) (High,87.5)};

    \addplot+[fill=lightgray, draw=black, bar shift=-3pt, area legend]
      coordinates {(Low,35.0) (Medium,60.0) (High,85.0)};

    \addplot+[pattern=dots, pattern color=black, draw=black, bar shift=3pt, area legend]
      coordinates {(Low,30.0) (Medium,66.67) (High,85.0)};

    \addplot+[pattern=crosshatch, pattern color=black, fill=white, draw=black, bar shift=9pt, area legend]
      coordinates {(Low,15.0) (Medium,46.67) (High,85.0)};

    \addplot+[fill=darkgray, draw=black, bar shift=15pt, area legend]
      coordinates {(Low,21.25) (Medium,61.67) (High,86.25)};

    \legend{GPT-3.5, o3, GPT-4o, 4o-mini, 4o-mini (ft), Avg}
    \end{axis}
  \end{tikzpicture}

\vspace{-1.5em}  \caption{Accuracy across retrieval-context quality (Low/Medium/High) for each backend paired with \textsc{\scheme{}}.  It shows that the retrieval quality is the precondition for cache replacement policy high level reasoning.  }
  \Description{}
  \label{fig:retrieval_context_accuracy}
\end{figure}

\vspace{-0.5em}
\paragraph{Trick Questions.}
Epistemic checks (5 queries) separate robust models: GPT-4o and GPT-4o-mini each reach 80.0\%, o3 20.0\%, and GPT-3.5 0.0\%. Example: \emph{``Does PC 0x4037aa in \texttt{lbm} access 0x1b73be82e3f under PARROT?''} is invalid because that PC appears only in \texttt{mcf}; models must reject the false premise rather than guess.

\vspace{-0.5em}
\paragraph{Simple Reasoning.}
On short, concept-style microarchitecture questions (5 items, scored 0--5 each), GPT-4o reaches 80.0\% (20/25), GPT-4o-mini 76.0\%, GPT-3.5 56.0\%, and o3 52.0\%. These do not strictly need retrieval but reward stable domain knowledge.

\vspace{-0.5em}
\paragraph{Code Generation.}
For retrieval-and-analysis code (5 items, 0--5 scoring), GPT-4o attains 100.0\% (25/25) and GPT-4o-mini 96.0\%. GPT-3.5 is surprisingly strong at 92.0\%, while o3 and finetuned 4o-mini trail at 52.0\% and 68.0\%. Clear schemas and narrow APIs appear to favor code synthesis.

\vspace{-0.5em}
\paragraph{Replacement Policy Analysis.}
Causal questions (5 items, 0--5) such as \emph{``Why does Belady outperform LRU on PC  0x409270 in \texttt{astar}?''} require linking reuse patterns to policy rules. GPT-4o leads at 84.0\%, followed by GPT-4o-mini 76.0\%, finetuned 4o-mini 72.0\%, and o3 60.0\%; GPT-3.5 is at 56.0\%. Larger models better connect retrieved evidence to policy logic.

\vspace{-0.5em}
\paragraph{Workload Analysis.}
Whole-workload behavior (5 items, 0--5) shows GPT-4o at 88.0\%, GPT-4o-mini 76.0\%, finetuned 4o-mini 68.0\%, while GPT-3.5 and o3 both reach 48.0\%. Summarizing patterns across many PCs remains sensitive to retrieval coverage.
\vspace{-0.5em}
\paragraph{Semantic Analysis.}
Linking trace events to code intent (5 items, 0--5) is hardest: GPT-4o-mini attains 76.0\%, GPT-4o 72.0\%, finetuned 4o-mini 48.0\%, o3 40.0\%, and GPT-3.5 28.0\%. Models must connect PC-level locality with disassembly/function context, going beyond symbolic lookups.



\begin{figure}[htbp]
  \centering
  \begin{tikzpicture}[
    node distance=6mm,
    box/.style={draw, rounded corners, align=left, inner sep=6pt, text width=0.95\linewidth},
    flowarrow/.style={-Stealth, thick}
  ]

  \node[box, fill=gray!10] (cat) {\textbf{Question Category:} Cache Hit/Miss};

  \node[box, below=of cat] (query) {%
    \textbf{User Query:}\\
    \texttt{Does the memory access with PC 0x401dc9 and address 0x47ea85d37f result in a cache hit or cache miss for the lbm workload and PARROT replacement policy?}
  };

  \node[box, below=of query] (fewshot) {%
    \textbf{One-shot Example provided to the Generator LLM via prompt:}\\[0.25em]
    \textbf{Context:}\\
    \ttfamily
    For policy LRU on workload lbm \dots{} at PC 0x401dc9 and address 0x47ea85d37f:\\
    Cache result: Cache Miss\\
    Evicted address: 0x19e02d19b7f (needed again in 2304 accesses), Inserted address needed again in 3132 accesses.\\
    Answer the following question: Does the memory access with PC 0x401dc9 and address 0x47ea85d37f result in a cache hit or cache miss for the lbm workload and LRU replacement policy? The correct answer is,\\
    \rmfamily
    \textbf{Response:} \texttt{Cache Miss}
  };

  \node[box, fill=gray!10, below=of fewshot] (out) {%
    \textbf{LLM Output:} \texttt{Cache Hit} or \texttt{Cache Miss}, grounded in the retrieved trace context.
  };

  \draw[flowarrow] (cat) -- (query);
  \draw[flowarrow] (query) -- (fewshot);
  \draw[flowarrow] (fewshot) -- (out);

  \end{tikzpicture}
\vspace{-1em}  \caption{A one-shot prompt example passed along with a Cache Hit/Miss user query  to the generator LLM and the correct \textsc{\scheme{}}'s output. } 
  \Description{Flow diagram with four stacked boxes: category, user query, one-shot example showing a prompt/response for LRU on lbm, and the expected LLM output form (Cache Hit or Cache Miss). Arrows connect top to bottom.}
  \label{fig:hitmiss_fewshot}
\end{figure}
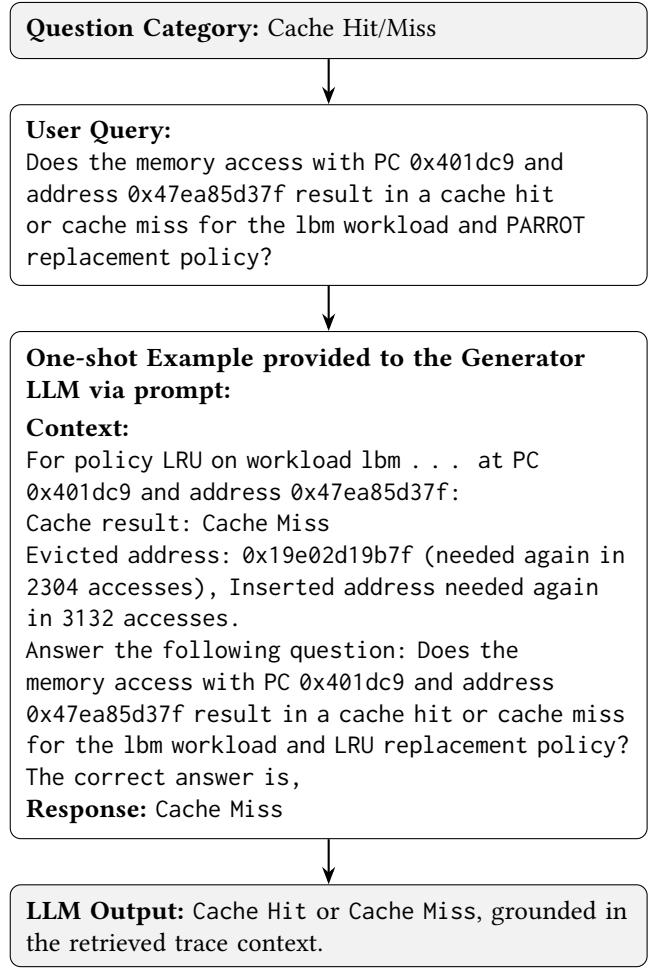

\begin{figure}[!htbp]
  \centering

  \begin{subfigure}{0.48\linewidth}
    \centering
    \begin{tikzpicture}
      \begin{axis}[
        ybar,
        ymin=0, ymax=12,
        ytick={0,2,4,6,8,10},
        ymajorgrids,
        y grid style={densely dashed, gray!50},
        xlabel={Score}, ylabel={Number of Questions},
        ylabel style={font=\footnotesize},
        xlabel style={font=\footnotesize},
        xtick={0,1,2,3,4,5},
        bar width=8pt,
        width=\linewidth, height=0.75\linewidth
      ]
        \addplot+[ybar, fill=lightgray, draw=black] coordinates {(0,0) (1,5) (2,4) (3,8) (4,2) (5,4)};
      \end{axis}
    \end{tikzpicture}
    \caption{GPT-3.5-Turbo}
  \end{subfigure}\hfill
  \begin{subfigure}{0.48\linewidth}
    \centering
    \begin{tikzpicture}
      \begin{axis}[
        ybar,
        ymin=0, ymax=12,
        ytick={0,2,4,6,8,10},
        ymajorgrids,
        y grid style={densely dashed, gray!50},
        xlabel={Score}, ylabel={Number of Questions},
        ylabel style={font=\footnotesize},
        xlabel style={font=\footnotesize},
        xtick={0,1,2,3,4,5},
        bar width=8pt,
        width=\linewidth, height=0.75\linewidth
      ]
        \addplot+[ybar, fill=lightgray, draw=black] coordinates {(0,0) (1,0) (2,1) (3,4) (4,8) (5,11)};
      \end{axis}
    \end{tikzpicture}
    \caption{GPT-4o}
  \end{subfigure}

  \begin{subfigure}{0.48\linewidth}
    \centering
    \begin{tikzpicture}
      \begin{axis}[
        ybar,
        ymin=0, ymax=12,
        ytick={0,2,4,6,8,10},
        ymajorgrids,
        y grid style={densely dashed, gray!50},
        xlabel={Score}, ylabel={Number of Questions},
        ylabel style={font=\footnotesize},
        xlabel style={font=\footnotesize},
        xtick={0,1,2,3,4,5},
        bar width=8pt,
        width=\linewidth, height=0.75\linewidth
      ]
        \addplot+[ybar, fill=lightgray, draw=black] coordinates {(0,9) (2,4) (3,2) (4,1) (5,9)};
      \end{axis}
    \end{tikzpicture}
    \caption{o3}
  \end{subfigure}\hfill
  \begin{subfigure}{0.48\linewidth}
    \centering
    \begin{tikzpicture}
      \begin{axis}[
        ybar,
        ymin=0, ymax=12,
        ytick={0,2,4,6,8,10},
        ymajorgrids,
        y grid style={densely dashed, gray!50},
        xlabel={Score}, ylabel={Number of Questions},
        ylabel style={font=\footnotesize},
        xlabel style={font=\footnotesize},
        xtick={0,1,2,3,4,5},
        bar width=8pt,
        width=\linewidth, height=0.75\linewidth
      ]
        \addplot+[ybar, fill=lightgray, draw=black] coordinates {(0,1) (2,7) (3,5) (4,6) (5,4)};
      \end{axis}
    \end{tikzpicture}
    \caption{Finetuned 4o-mini}
  \end{subfigure}

  \caption{Distribution of reasoning scores by model. Each panel shows the frequency of scores \{0,\dots,5\} for the corresponding backend when paired with \textsc{\scheme{}}: showing how many questions each model scored from 0 to 5}.
  \Description{Four bar charts showing how many questions each model scored at levels 0 through 5. GPT-4o concentrates on higher scores; o3 is bimodal with many 0 and 5; GPT-3.5 clusters around middle scores; finetuned 4o-mini sits between.}
  \label{fig:model_scores}
  \vspace{-1em}
\end{figure}
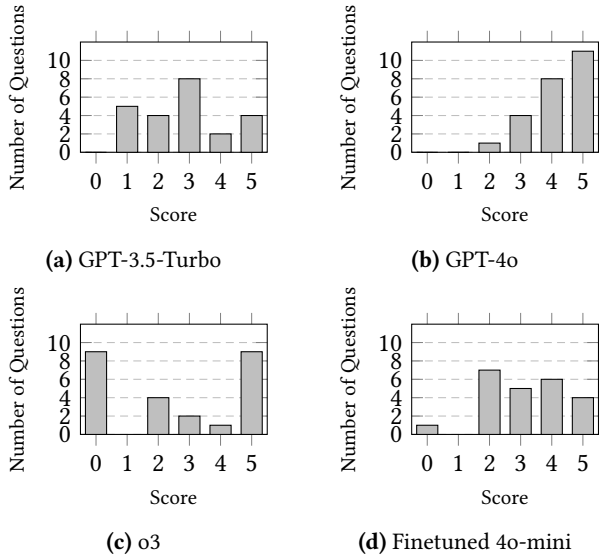


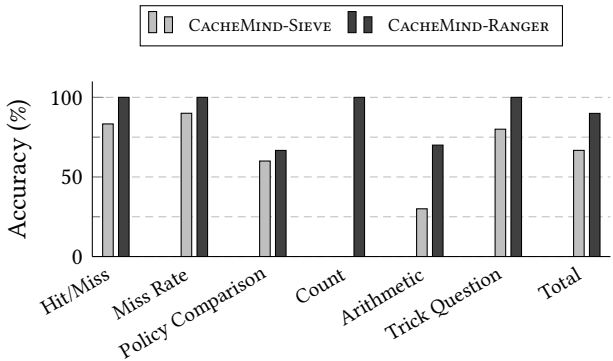
\begin{figure}[!htbp]
\centering
\begin{tikzpicture}
\begin{axis}[
    ybar,
    bar width=4pt,
    width=\linewidth,
    height=0.22\textwidth,
    ymin=0,
    ymax=110,
    ylabel={Accuracy (\%)},
    symbolic x coords={
        Hit/Miss, Miss Rate, Policy Comparison, Count, Arithmetic, Trick Question, Total
    },
    xtick=data,
    xtick style={draw=none},
    x tick label style={rotate=30, anchor=east, font=\footnotesize},
    y tick label style={font=\small},
    axis x line*=bottom,
    axis y line*=left,
    enlarge x limits=0.05,
    ymajorgrids,
    yminorgrids,
    minor y tick num=1,
    y grid style={densely dashed, gray!50},
    minor y grid style={densely dashed, gray!50},
    legend style={
        at={(0.5,1.2)}, anchor=south,
        legend columns=5,
        column sep=4pt,
        font=\scriptsize
    }
]

\addplot+[fill=lightgray, draw=black] coordinates {
    (Hit/Miss,83.3) (Miss Rate,90) (Policy Comparison,60)
    (Count,0) (Arithmetic,30) (Trick Question,80) 
    (Total,66.7)
};

\addplot+[fill=darkgray, draw=black] coordinates {
    (Hit/Miss,100) (Miss Rate,100) (Policy Comparison,66.67)
    (Count,100) (Arithmetic,70) (Trick Question,100) (Total,89.93)
};

\legend{
    \textsc{\scheme{}-Sieve},
    \textsc{\scheme{}-Ranger},
}

\end{axis}
\end{tikzpicture}
\vspace{-1em}\caption{Accuracy of the two proposed Retrievers - \textsc{\scheme{}-Sieve} and \textsc{\scheme{}-Ranger} across \textsc{\scheme{}Bench} categories. \textsc{\scheme{}-Ranger} achieves an average score of 89.33\% compared to 66.67\% of \textsc{\scheme{}-Sieve} on Trace-Grounded Questions. This highlights the potential of dynamic LLM-based retrievers.}
\Description{}
\label{fig:retriever_comparison}
\end{figure}

\begin{figure*}[!htbp]
\centering
\begin{tikzpicture}
  \def\W{\textwidth}
  \def\Hh{0.7cm}
  \def\Hq{0.9cm}      
  \def\Hr{8cm}      
  \def\He{2.25cm}      
  \def\Pad{6pt}

  \coordinate (TL) at (0,0);
  \coordinate (BR) at (\W,-\Hh-\Hq-\Hr-\He); 

  \draw[thick] (TL) rectangle (BR);
  \draw[thick] (0,-\Hh) -- (\W,-\Hh);
  \draw[thick] (0,-\Hh-\Hq) -- (\W,-\Hh-\Hq);
  \draw[thick] (0,-\Hh-\Hq-\Hr) -- (\W,-\Hh-\Hq-\Hr); 

  \draw[thick] (\W/3,0) -- (\W/3,-\Hh);
  \draw[thick] (2*\W/3,0) -- (2*\W/3,-\Hh);

  \draw[thick] (\W/3,-\Hh-\Hq) -- (\W/3,-\Hh-\Hq-\Hr);
  \draw[thick] (2*\W/3,-\Hh-\Hq) -- (2*\W/3,-\Hh-\Hq-\Hr);

  \draw[thick] (\W/3,-\Hh-\Hq-\Hr) -- (\W/3,-\Hh-\Hq-\Hr-\He); 
  \draw[thick] (2*\W/3,-\Hh-\Hq-\Hr) -- (2*\W/3,-\Hh-\Hq-\Hr-\He); 

  \node[font=\bfseries, align=center] at (\W/6,   -\Hh/2) {LlamaIndex};
  \node[font=\bfseries, align=center] at (\W/2,   -\Hh/2) {\textsc{Sieve}};
  \node[font=\bfseries, align=center] at (5*\W/6, -\Hh/2) {\textsc{Ranger}};

  \node[anchor=west, font=\bfseries] at (\Pad,-\Hh-\Pad) {Example Query:};
  \node[
    anchor=north west,
    align=left,
    text width=\dimexpr\W-2\Pad\relax
  ] at (\Pad,-\Hh-0.35cm) {%
    \small\emph{When PC \texttt{0x409270} and address \texttt{0x2bfd401c63f} is accessed on the astar workload with LRU policy, does the cache hit or miss?}%
  };

  \node[anchor=west, font=\bfseries] at (\Pad,-\Hh-\Hq-\Pad) {Example Retrieved Context:};

  \def\CW{\dimexpr\W/3-2\Pad\relax}

  \node[
    anchor=north west,
    align=left,
    text width=\CW
  ] at (\Pad,-\Hh-\Hq-0.35cm) {%
    \footnotesize\ttfamily
    \raggedright
    \sloppy
    0.6483328952344215 \\
    \{`trace\_id': `astar\_evictions\_lru', `doc\_type': `trace\_summary'\} \\
    DESCRIPTION: Replacement Policy: <Policy Description>\\
    Workload: <Workload Description> \\
    --- \\
    0.6140775745472963 \\
    TRACE\_ID: astar\_evictions\_lru \\
    \textbf{program\_counter=0x409538, memory\_address=0x2bfd401b693,} evict=Cache Miss,\\
    ---\\
    0.6140775745472963 \\
    TRACE\_ID: astar\_evictions\_lru \\
    \textbf{program\_counter=0x4090c3, memory\_address=0x2bfd401caf2,} evict=Cache Miss,\\
    
    \ldots \\
    \textbf{\textit{Exact PC, Memory Address match not found}}
  };

\node[
  anchor=north west,
  align=left,
  text width=\CW
] at (\W/3+\Pad,-\Hh-\Hq-0.35cm) {%
  \footnotesize\ttfamily\raggedright\sloppy
  Workloads involved:\\
  <Workload Description>\\[2pt]
  Policies involved:\\
  <Policy Description>\\[2pt]
  Assembly code snippet for PC \texttt{0x409270} (representative instructions):\\
  <Assembly Code>\\[2pt]
  Source Function: \texttt{\_ZN7way2obj11createwayarERP6pointtRi}\\[4pt]
  \textbf{LRU + astar @ PC \texttt{0x409270}, addr \texttt{0x2bfd401c63f}:} \\
  \textbf{Cache result: Cache Miss}\\[2pt]
  Belady + astar: \ldots\\
  MLP + astar: \ldots\\[2pt]
  Evicted address: \texttt{0x31232a3ee3f}
  (needed again in 3648 accesses).\\
  Inserted address needed again in 6287 accesses.%
};

\node[
  anchor=north west,
  align=left,
  text width=\CW
] at (2*\W/3+\Pad,-\Hh-\Hq-0.35cm) {%
  \footnotesize\ttfamily\raggedright\sloppy
  \textbf{Result: Cache Miss for PC \texttt{0x409270} and addr \texttt{0x2bfd401c63f}} \\
  (workload: astar, policy: LRU).\\[2pt]
  Function: \texttt{\_ZN7way2obj11createwayarERP6pointtRi}\\
  Assembly: <Assembly Code>\\[2pt]
  Cache Performance Summary:\\
  140704 total accesses; 133542 total misses; 94.91\% miss rate; \ldots\\[2pt]
  Replacement Policy:\\
  <Policy Description>\\
  Workload: <Workload Description>%
};

  \node[anchor=west, font=\bfseries] at (\Pad,-\Hh-\Hq-\Hr-\Pad) {Evaluation:};

  \node[
    anchor=north west,
    align=left,
    text width=\CW
  ] at (\Pad,        -\Hh-\Hq-\Hr-\Pad-0.25cm) 
  {
  \small
  \textit{Queries with correct retrieved context = \textbf{1/10 (10\%)} \\
  Average time for retrieval = \textbf{36.62 seconds}}
  };
  
  \node[
    anchor=north west,
    align=left,
    text width=\CW
  ] at (\W/3+\Pad,   -\Hh-\Hq-\Hr-\Pad-0.25cm) 
  {
  \small
  \textit{Queries with correct retrieved context = \textbf{6/10 (60\%)} \\
  Average time for retrieval = \textbf{3.65 seconds}}
  };
    
  \node[
    anchor=north west,
    align=left,
    text width=\CW
  ] at (2*\W/3+\Pad, -\Hh-\Hq-\Hr-\Pad-0.25cm) 
  {
  \small 
  \textit{Queries with correct retrieved context = \textbf{9/10 (90\%)} \\
  Average time for retrieval = \textbf{4.43 seconds}}
  };

\end{tikzpicture}
\vspace{-1.5em}
\caption{Comparison of retrieved context across LlamaIndex, \textsc{Sieve}, and \textsc{Ranger} for the same query.}
\label{fig:retrieval_comparison}
\end{figure*}

\vspace{1em}
\noindent\textbf{Finetuning.} 
Contrary to expectation, Figure~\ref{fig:main} shows that fine-tuned GPT-4o-mini did not outperform the non-fine-tuned variant. In reasoning categories (Trick Question, Semantic Analysis), fine-tuning amplified hallucinations (20\% vs. 80\% and 48\% vs. 76\% respectively). This aligns with recent findings~\cite{gekhman2024-fine} showing that fine-tuning narrows generalization and increases hallucination risk in scientific domains. 

\noindent\textbf{One and Few-shot Prompting. }
Figure ~\ref{fig:hitmiss_fewshot} shows an example of the \textsc{\scheme{}} one-shot prompt. 
Based on our results, we observed that overall, one or few-shot prompting does not improve system performance significantly. For few questions, when the generator LLM is not provided with sufficient context, it takes the context from the example as its own and gives incorrect answers. On the other hand, the given examples help the generator identify and assess trick questions better than zero-shot prompting.

\noindent\textbf{LLM Variation. \label{sec_6_5}}
Across all 11 benchmark categories, GPT-4o paired with \textsc{\scheme{}} achieves the highest weighted total score of 74.9\%, followed by o3 (64.8\%), finetuned 4o-mini (62.7\%), and GPT-3.5 (60.0\%). GPT-3.5 serves as a weak but useful baseline, while o3 shows strong reasoning but inconsistent response coverage (Figure~\ref{fig:model_scores}). 

\subsection{\textsc{\scheme{}} vs.  Conventional Retrieval/RAGs}\label{sec_6_rag}

\textsc{Sieve} exemplifies a conventional retrieval approach, it performs poorly even on primitive tasks (Figure ~\ref{fig:retriever_comparison}), confirming that a static program cannot generalize to the diverse, arbitrary decomposition-based questions \textsc{\scheme{}} is designed to handle. 
Figure~\ref{fig:retrieval_comparison} compares retrieval accuracy across ten evaluation queries spanning five categories of trace-grounded questions.
LlamaIndex achieves 10\% accuracy, \textsc{Sieve} achieves 60\%, and \textsc{Ranger} achieves 90\%.
These results show that \textsc{\scheme{}} significantly improves retrieval accuracy for non-trivial natural-language queries over microarchitectural traces.

This failing of prior RAGs 
stems from fundamental limitations of embedding-based RAGs.
Conventional frameworks such as LlamaIndex~\cite{llamaindex} rely on c\textit{osine similarity} over embeddings, which works well for text-heavy datasets but fails for microarchitectural traces where records differ only by small numerical or bit-level changes.
As a result, embedding-based retrievers often return imprecise or incorrect context.
To address this issue, we first designed \textsc{\scheme{}-Sieve}, a filter-based retriever that performs explicit checks on structured attributes such as program counter (PC), memory address, replacement policy, and workload.
While \textsc{Sieve} improves accuracy, it requires anticipating query patterns in advance, limiting its flexibility similar to conventional database retrievals.

\textsc{\scheme{}-Ranger} removes this limitation by using the code-generation capability of large language models to dynamically generate database queries at inference time, enabling accurate retrieval for previously unseen query formulations.

\subsection{Actionable Insights} \label{sec:actionable}

\textcolor{black}{This section provides  concrete actionable insights 
derived using \textsc{\scheme{}} 
showing the utility of \textsc{\scheme{}} on non-trivial queries that require a natural-language interface directly providing speedup.}

\begin{figure}[!htbp]
\centering
\newlength{\ChatW}
\setlength{\ChatW}{\linewidth}
\begin{tikzpicture}
  \def\H{6.25cm}  
  \def\Pad{6pt}

  \draw[thick] (0,0) rectangle (\ChatW,-\H);

  \node[
    anchor=north west,
    align=left,
    text width=\dimexpr\ChatW-2\Pad\relax
  ] at (\Pad,-\Pad) {%
    \small
    \textbf{User:} Mockingjay uses PC-based reuse-distance prediction; suggest ideas to improve performance.\\
    \textbf{Assistant:} \textit{Summary of ideation: cluster PCs by ETR \\variance; train RDP on stable samples.}\\[4pt]

    \textbf{User:} List all unique PCs in the trace.\\
    \textbf{Assistant:} \texttt{4184b0, 4184c0, 418502, \ldots, 417f58}.\\[4pt]

    \textbf{User:} Compute mean ETR per PC.\\
    \textbf{Assistant:} \texttt{PC\_i $\rightarrow$ mean\_ETR(PC\_i)} (collapsed).\\[4pt]

    \textbf{User:} Compute standard deviation of ETR per PC.\\
    \textbf{Assistant:} \texttt{PC\_i $\rightarrow$ std\_ETR(PC\_i)} (collapsed).\\[4pt]

    \textbf{User:} Group PCs by ETR variance.\\
    \textbf{Assistant:} LowVar: \texttt{\{413930, \ldots\}}, 
    MedVar: \texttt{\{41391c, \ldots\}}, 
    HighVar: \texttt{\{413948, \ldots\}}.%
  };
\end{tikzpicture}
\vspace{-0.4em}
\caption{Identifying PCs with low ETR variance (more predictable reuse) using \texttt{\textsc{\scheme{}}} for enhancing Mockingjay.}
\label{fig:chat_session_mockingjay}
\end{figure}


\noindent\textbf{Mockingjay Use Case.} 
\textcolor{black}{
The Mockingjay \cite{mockingjay} replacement policy approximates Belady's Optimal (MIN) policy by predicting the reuse distance of cache lines in accessed cache set and evicting the line with the furthest reuse distance in the future. It uses a multiclass PC-based reuse distance predictor and orders cache lines by their estimated time of reuse (ETR). 
Figure \ref{fig:chat_session_mockingjay} shows the chat session using which we can identify PCs with different amounts of variance in their ETR value, which in turn helps us identify PCs that have more predictable reuse distance values. We can use these "stable" PCs to train the Reuse Distance Predictor with less noisy samples.  
}
\textcolor{black}{
Based on the above insight, we changed the Mockingjay source code to train only on the list of stable PCs identified by \textsc{\scheme{}}. We compared the results of both modified and unmodified policies on ChampSim, keeping everything else the same, to observe that stable training increased IPC from 0.47698 to 0.480307 (0.7\% speedup) over milc workload. We chose milc because Mockingjay does worse than Hawkeye. 
}

\vspace{0.5em}
\noindent\textbf{Signature Optimization for Bypass Logic.} 
\textcolor{black}{
On the \textit{mcf} workload, bypassing the PCs identified by \textsc{\scheme{}} improves performance under an LRU policy.
Specifically, bypassing ten PCs increases the cache hit rate from 25.06\% to 26.98\% (+7.66\% relative) and improves IPC from 0.047905 to 0.048809, corresponding to a 2.04\% speedup.
All other simulation parameters are kept unchanged.
}

\begin{figure}[!htbp]
\centering
\setlength{\ChatW}{\linewidth}

\begin{tikzpicture}
  \def\H{11.5cm}   
  \def\Pad{5pt}

  \draw[thick] (0,0) rectangle (\ChatW,-\H);

  \node[
    anchor=north west,
    align=left,
    text width=\dimexpr\ChatW-2\Pad\relax
  ] at (\Pad,-\Pad) {%
    \small
    \textbf{User:} List all PCs in the mcf workload.\\
    \textbf{Assistant:} \texttt{0x401380, 0x401384, 0x40138f, \ldots}\\[3pt]

    \textbf{User:} For mcf + Belady, compute average accessed-address reuse distance, cache hit rate and hit count per PC; sort in descending order in terms of reuse distance.\\
    \textbf{Assistant:} \texttt{PC $\rightarrow$ (AvgReuseDist, HitRate, HitCount)} \\ (e.g., \texttt{0x402ea8}, \texttt{0x4037aa}, \ldots).\\[3pt]


    \textbf{User:} Identify PCs suitable for bypassing to improve IPC.\\
    \textbf{Assistant:} Bypass candidates:
    \texttt{0x4037aa, 0x402ea8, 0x401380, 0x40138f, \ldots}\\
    (high reuse distance and/or near-zero hit rate).\\[2pt]
    \texttt{\ldots}\\[2pt]

    \textbf{User:} Based on the reuse and hit statistics for lbm workload, what insights can you derive?\\
    \textbf{Assistant:} In lbm, interleaved streaming accesses appear to push useful lines to LRU positions long before reuse. 
    Implications and how to approximate Belady for lbm: \\
    \begin{itemize}
        \item Use PC-aware reuse prediction. 
        \item Favor RRIP/DIP-like insertion/aging.
        \item Consider bypass and scan detection.
    \end{itemize}
    Net insight: lbm’s working set interleaves strong temporal reuse with heavy scan interference. Belady’s advantage comes from separating these classes at fine granularity and preserving the long-horizon reuses that recency- and heuristic-driven policies fail to capture.
  };
\end{tikzpicture}

\vspace{-0.4em}
\caption{Condensed transcript of an LLM-assisted analysis for identifying bypassable PCs in the \texttt{mcf} workload. Long numerical outputs are collapsed for clarity.}
\label{fig:chat_bypass}
\end{figure}

\textcolor{black}{
Figure~\ref{fig:chat_bypass} shows the \textsc{\scheme{}} queries used to identify PCs in the \textit{astar}, \textit{lbm}, and \textit{mcf} traces that exhibit low hit rates and large reuse distances under Belady’s optimal policy.
Such PCs are frequently evicted even by the optimal policy and compete with other PCs that demonstrate higher reuse.
This suggests that bypassing these accesses can reduce cache pollution.
}

\textcolor{black}{
To validate this insight, we ran ChampSim simulations using the same workloads and simpoints used to generate the \textsc{\scheme{}} traces.
We implemented a simple conditional bypass in the LRU replacement logic that skips cache insertion for the identified PCs.
}
\textcolor{black}{
\textsc{\scheme{}} also reveals complex access behavior in the \textit{lbm} workload, where PCs with high reuse are interleaved with PCs exhibiting poor reuse, as shown in Figure ~\ref{fig:chat_bypass}.
This observation helps understand why PC-signature based policies such as SHiP~\cite{ship} outperform their predecessor policies like RRIP ~\cite{rrip} on \textit{lbm}, as they learn reuse behavior directly at PC granularity.
}

\vspace{0.5em}
\noindent\textbf{PC-Information Applications, Software Intervention and Prefetcher Use Case.} 
\textcolor{black}{
Using the PC identified by \textsc{\scheme{}}, adding a software prefetch to a pointer-chasing microbenchmark increases IPC from 0.131452 to 0.231261.
}
\textcolor{black}{
Figure~\ref{fig:chat_prefetcher} shows the \textsc{\scheme{}} queries used to identify the PC responsible for the majority of cache misses in the microbenchmark.
The benchmark is designed to generate misses from a single dominant load instruction at an initially unknown PC, which is recovered using \textsc{\scheme{}}. 
}
\textcolor{black}{
Based on this insight, we modified the microbenchmark to insert a built-in C software prefetch instruction that prefetches future addresses in the pointer-chasing array according to the observed access pattern, leading to significant performance gains.
Simulating this modified source code 
demonstrates that \textsc{\scheme{}} can identify actionable PCs and directly guide performance-improving software-hardware co-design, intervention and fixes. 
}

\begin{figure}[!htbp]
\centering
\setlength{\ChatW}{\linewidth}

\begin{tikzpicture}
  \def\H{3.25cm}   
  \def\Pad{5pt}

  \draw[thick] (0,0) rectangle (\ChatW,-\H);

  \node[
    anchor=north west,
    align=left,
    text width=\dimexpr\ChatW-2\Pad\relax
  ] at (\Pad,-\Pad) {%
    \small
    \textbf{User:} List all unique PCs in the given trace.\\
    \textbf{Assistant:} \texttt{0x400444, 0x400701, 0x400709, \ldots} 
    \\[4pt]

    \textbf{User:} From the unique PCs, identify the PC causing the most cache misses.\\
    \textbf{Assistant:} \texttt{0x400512}.\\[4pt]

    \textbf{User:} What is the miss rate of PC \texttt{0x400512}?\\
    \textbf{Assistant:} \textbf{74.73\%} miss rate.\\[4pt]
  };
\end{tikzpicture}

\vspace{-0.4em}
\caption{Condensed transcript of an interactive analysis identifying the most miss-prone PC in a cache access trace. Long PC lists are collapsed for clarity.}
\label{fig:chat_prefetcher}
\end{figure}

 \vspace{0.5em}
\noindent\textbf{Set Hotness Analysis Use Case.} 
\textcolor{black}{
Figure~\ref{fig:chat_session_cache_sets} shows how \textsc{\scheme{}} is used to identify hot and cold cache sets from access traces.
This analysis allows users to quickly localize sets that dominate cache activity, which is difficult to infer directly from raw traces.}
\textcolor{black}{
Knowledge of set hotness is important for understanding cache behavior and replacement effectiveness.
In sampled-set LLC policies, such as those used in prior work, learning eviction behavior from hot sets is more effective than uniform random sampling, since hot sets generate more misses and replacement decisions.
}

\textcolor{black}{
By providing a natural-language interface to query set-level activity, \textsc{\scheme{}} supports targeted analysis of cache behavior and enables more informed sampling and replacement policy design.
}

\begin{figure}[!htbp]
\centering
\begin{tikzpicture}
  \def\W{\linewidth}
  \def\H{11.5cm}   
  \def\Pad{6pt}

  \draw[thick] (0,0) rectangle (\W,-\H);

  \node[
    anchor=north west,
    align=left,
    text width=\dimexpr\W-2\Pad\relax
  ] at (\Pad,-\Pad) {%
    \small
    \textbf{User:} For astar workload and Belady replacement policy, could you list unique cache sets in ascending order?\\
    \textbf{Assistant:} \texttt{6, 35, 38, 53, 67, \ldots, 1940, 1967, 1973}.\\[5pt]

    \textbf{User:} Based on the list of cache sets accessed for astar + Belady, could you find the number of cache hits and hit rate?\\
    \textbf{Assistant:} \texttt{Set S\_i: hits(S\_i), hit rate(S\_i)} for $i=1\ldots K$ (full per-set table).\\
    e.g., Set 6: 863 hits, 38.32\%.\\[5pt]

    \textbf{User:} Identify 5 hot and 5 cold sets by hit rate.\\
    \textbf{Assistant:} \texttt{Hot Sets = \{332, 1424, 1604, 1086, 1456\}, Cold Sets = \{35, 53, 143, 362, 1211\}}.\\[5pt]

    \textbf{User:} For astar workload and LRU replacement policy, could you list unique cache sets in ascending order?\\
    \texttt{\ldots}\\[5pt]

    \textbf{Assistant:} \texttt{Hot Sets = \{332, 1604, 1424, 1456, 1086\}, Cold Sets = \{287, 1789, 113, 157, 488\}}.\\[5pt]

    \textbf{User:} Compare hot sets (LRU vs Belady) and derive insights.\\
    \textbf{Assistant:} Hot set identity likely overlaps, but Belady amplifies hotness:
    \begin{itemize}\itemsep0pt
      \item Hot sets arise from intrinsic workload locality.
      \item Belady avoids premature evictions, increasing hit concentration.
    \end{itemize}
  };
\end{tikzpicture}

\vspace{-0.3em}
\caption{Demo chat session on \textsc{\scheme{}} for identifying hot and cold cache sets and deriving insights.}
\label{fig:chat_session_cache_sets}
\end{figure}

Beyond aggregate accuracy, the experiments reveal deeper findings:

   
   \noindent\textbf{Belady vs. PARROT.} Across the three benchmarks, astar, lbm, and mcf, PARROT outperformed Belady for 2, 5, and 3 PCs \textcolor{black}{respectively, in terms of hit rate}. Although this may seem counterintuitive, it highlights a key distinction: Belady (OPT) makes globally optimal decisions over the entire trace, whereas PARROT learns PC-local heuristics. Consequently, OPT exploits structure beyond PC identity that PARROT and other PC-based policies fail to capture. \textcolor{black}{OPT provides an upper bound on the total cache hit rate, implying that any practical replacement policy achieves a hit rate less than or equal to OPT when aggregated over all accesses. However, this global guarantee does not extend to individual program counters (PCs): the hit rate of OPT for accesses associated with a given PC may be lower than that of a practical replacement policy, as observed.}
   
 \vspace{0.5em}
 \noindent \textbf{Distribution of Model Behavior.} Score distributions (Figure~\ref{fig:model_scores}) reveal that o3 is bimodal, excelling or failing completely, while GPT-4o shows consistent competence across categories. GPT-3.5-Turbo and Finetuned GPT-4o-mini also show uniform distribution with a lower mean.
 
\vspace{0.5em}
 \noindent \textbf{Context can suppress latent knowledge.} In a \emph{Microarchitecture Concepts} question that required decomposing a memory address into \emph{offset/index/tag} given the cache geometry, all Generator models answered incorrectly when paired with \textsc{\scheme{}-Sieve} (which surfaced partial/ambiguous metadata). When paired with \textsc{\scheme{}-Ranger}, which retrieved the correct line size and set count, the same models produced the correct decomposition. 
    This suggests the LLMs possess the requisite background knowledge but are highly sensitive to retrieval precision: inappropriate context can override otherwise correct internal reasoning.
    
\vspace{0.5em}
 \noindent \textbf{Trust and Epistemic Robustness.} Trick Question performance suggests that only GPT-4o achieves the rejection behavior necessary for deployment in hardware verification workflows. Other models hallucinate under adversarial phrasing.
 
\vspace{0.5em}
\noindent \textbf{Semantic Recovery.} Linking low-level cache events to high-level code structures remains the hardest benchmark. This opens new opportunities for applying \textsc{\scheme{}} as a co-analysis tool in architecture research (Figure~\ref{fig:trace_data} shows an example of such capability).


\section{Conclusion} \label{sec_7}
\textsc{\scheme{}} reframes cache analysis as \emph{query-first} reasoning over traces, coupling a symbolic–semantic retriever with LLMs to turn raw events into verifiable answers. 
\textsc{\scheme{}} is a microarchitectural microscope that magnifies cache events to per-PC, per-data resolution. The  links each access to its program semantics and shows exactly which data each PC missed or evicted under every replacement policy. This fine-grained lens reveals patterns and behaviors that aggregate metrics obscure, enabling cross-policy comparisons and targeted design evaluation.
 
Across 100 verified questions, \textsc{\scheme{}} achieves strong accuracy, especially on hit/miss and miss-rate queries, and reveals that retrieval precision is the dominant factor, while large, reasoning-capable LLMs further lift performance.
Our LLM-based retriever (\textsc{\scheme{}-Ranger}) improves open-ended tasks and mitigates failures seen with a purely filter-driven pipeline (\textsc{\scheme{}-Sieve}). Fine-tuning offered limited benefit, whereas grounded context consistently reduced hallucinations and enabled policy/workload analysis.



\section*{Code and Data Availability}
The code for \textsc{\scheme{}} and the suite of benchmark questions, \textsc{\scheme{}Bench} are open-source and available at: \url{https://github.com/kaushal1803/cachemind}


\bibliographystyle{ACM-Reference-Format}
\bibliography{ref}


\begin{thebibliography}{39}


\ifx \showCODEN    \undefined \def \showCODEN     #1{\unskip}     \fi
\ifx \showISBNx    \undefined \def \showISBNx     #1{\unskip}     \fi
\ifx \showISBNxiii \undefined \def \showISBNxiii  #1{\unskip}     \fi
\ifx \showISSN     \undefined \def \showISSN      #1{\unskip}     \fi
\ifx \showLCCN     \undefined \def \showLCCN      #1{\unskip}     \fi
\ifx \shownote     \undefined \def \shownote      #1{#1}          \fi
\ifx \showarticletitle \undefined \def \showarticletitle #1{#1}   \fi
\ifx \showURL      \undefined \def \showURL       {\relax}        \fi
\providecommand\bibfield[2]{#2}
\providecommand\bibinfo[2]{#2}
\providecommand\natexlab[1]{#1}
\providecommand\showeprint[2][]{arXiv:#2}

\bibitem[Asai et~al\mbox{.}(2023)]%
        {selfrag}
\bibfield{author}{\bibinfo{person}{Akari Asai}, \bibinfo{person}{Zeqiu Wu}, \bibinfo{person}{Yizhong Wang}, \bibinfo{person}{Avirup Sil}, {and} \bibinfo{person}{Hannaneh Hajishirzi}.} \bibinfo{year}{2023}\natexlab{}.
\newblock \showarticletitle{Self-RAG: Learning to Retrieve, Generate, and Critique through Self-Reflection}.
\newblock  (\bibinfo{year}{2023}).
\newblock
\showeprint[arxiv]{2310.11511}~[cs.CL]
\urldef\tempurl%
\url{https://arxiv.org/abs/2310.11511}
\showURL{%
\tempurl}


\bibitem[Belady(1966)]%
        {belady}
\bibfield{author}{\bibinfo{person}{L.~A. Belady}.} \bibinfo{year}{1966}\natexlab{}.
\newblock \showarticletitle{A study of replacement algorithms for a virtual-storage computer}.
\newblock \bibinfo{journal}{\emph{IBM Systems Journal}} \bibinfo{volume}{5}, \bibinfo{number}{2} (\bibinfo{year}{1966}), \bibinfo{pages}{78--101}.
\newblock
\href{https://doi.org/10.1147/sj.52.0078}{doi:\nolinkurl{10.1147/sj.52.0078}}


\bibitem[Bera et~al\mbox{.}(2022)]%
        {hermes}
\bibfield{author}{\bibinfo{person}{Rahul Bera}, \bibinfo{person}{Konstantinos Kanellopoulos}, \bibinfo{person}{Shankar Balachandran}, \bibinfo{person}{David Novo}, \bibinfo{person}{Ataberk Olgun}, \bibinfo{person}{Mohammad Sadrosadati}, {and} \bibinfo{person}{Onur Mutlu}.} \bibinfo{year}{2022}\natexlab{}.
\newblock \showarticletitle{Hermes: Accelerating Long-Latency Load Requests via Perceptron-Based Off-Chip Load Prediction}. In \bibinfo{booktitle}{\emph{Proceedings of the 55th Annual IEEE/ACM International Symposium on Microarchitecture (MICRO)}} (Chicago, Illinois, USA). \bibinfo{publisher}{IEEE/ACM}, \bibinfo{pages}{1--18}.
\newblock
\href{https://doi.org/10.1109/MICRO56248.2022.00015}{doi:\nolinkurl{10.1109/MICRO56248.2022.00015}}


\bibitem[Borgeaud et~al\mbox{.}(2022)]%
        {retro}
\bibfield{author}{\bibinfo{person}{Sebastian Borgeaud}, \bibinfo{person}{Arthur Mensch}, \bibinfo{person}{Jordan Hoffmann}, \bibinfo{person}{Trevor Cai}, \bibinfo{person}{Eliza Rutherford}, \bibinfo{person}{Katie Millican}, \bibinfo{person}{George van~den Driessche}, \bibinfo{person}{Jean-Baptiste Lespiau}, \bibinfo{person}{Bogdan Damoc}, \bibinfo{person}{Aidan Clark}, \bibinfo{person}{Diego de Las~Casas}, \bibinfo{person}{Aurelia Guy}, \bibinfo{person}{Jacob Menick}, \bibinfo{person}{Roman Ring}, \bibinfo{person}{Tom Hennigan}, \bibinfo{person}{Saffron Huang}, \bibinfo{person}{Loren Maggiore}, \bibinfo{person}{Chris Jones}, \bibinfo{person}{Albin Cassirer}, \bibinfo{person}{Andy Brock}, \bibinfo{person}{Michela Paganini}, \bibinfo{person}{Geoffrey Irving}, \bibinfo{person}{Oriol Vinyals}, \bibinfo{person}{Simon Osindero}, \bibinfo{person}{Karen Simonyan}, \bibinfo{person}{Jack~W. Rae}, \bibinfo{person}{Erich Elsen}, {and} \bibinfo{person}{Laurent Sifre}.} \bibinfo{year}{2022}\natexlab{}.
\newblock \bibinfo{title}{Improving language models by retrieving from trillions of tokens}.
\newblock
\showeprint[arxiv]{2112.04426}~[cs.CL]
\urldef\tempurl%
\url{https://arxiv.org/abs/2112.04426}
\showURL{%
\tempurl}


\bibitem[Brown et~al\mbox{.}(2020)]%
        {gpt3}
\bibfield{author}{\bibinfo{person}{Tom~B. Brown}, \bibinfo{person}{Benjamin Mann}, \bibinfo{person}{Nick Ryder}, \bibinfo{person}{Melanie Subbiah}, \bibinfo{person}{Jared Kaplan}, \bibinfo{person}{Prafulla Dhariwal}, \bibinfo{person}{Arvind Neelakantan}, \bibinfo{person}{Pranav Shyam}, \bibinfo{person}{Girish Sastry}, \bibinfo{person}{Amanda Askell}, \bibinfo{person}{Sandhini Agarwal}, \bibinfo{person}{Ariel Herbert-Voss}, \bibinfo{person}{Gretchen Krueger}, \bibinfo{person}{Tom Henighan}, \bibinfo{person}{Rewon Child}, \bibinfo{person}{Aditya Ramesh}, \bibinfo{person}{Daniel~M. Ziegler}, \bibinfo{person}{Jeffrey Wu}, \bibinfo{person}{Clemens Winter}, \bibinfo{person}{Christopher Hesse}, \bibinfo{person}{Mark Chen}, \bibinfo{person}{Eric Sigler}, \bibinfo{person}{Mateusz Litwin}, \bibinfo{person}{Scott Gray}, \bibinfo{person}{Benjamin Chess}, \bibinfo{person}{Jack Clark}, \bibinfo{person}{Christopher Berner}, \bibinfo{person}{Sam McCandlish}, \bibinfo{person}{Alec Radford}, \bibinfo{person}{Ilya Sutskever},
  {and} \bibinfo{person}{Dario Amodei}.} \bibinfo{year}{2020}\natexlab{}.
\newblock \bibinfo{title}{Language Models are Few-Shot Learners}.
\newblock
\showeprint[arxiv]{2005.14165}~[cs.CL]
\urldef\tempurl%
\url{https://arxiv.org/abs/2005.14165}
\showURL{%
\tempurl}


\bibitem[Clark et~al\mbox{.}(2018)]%
        {arc}
\bibfield{author}{\bibinfo{person}{Peter Clark}, \bibinfo{person}{Isaac Cowhey}, \bibinfo{person}{Oren Etzioni}, \bibinfo{person}{Tushar Khot}, \bibinfo{person}{Ashish Sabharwal}, \bibinfo{person}{Carissa Schoenick}, {and} \bibinfo{person}{Oyvind Tafjord}.} \bibinfo{year}{2018}\natexlab{}.
\newblock \showarticletitle{Think you have Solved Question Answering? Try ARC, the AI2 Reasoning Challenge}.
\newblock  (\bibinfo{year}{2018}).
\newblock
\showeprint[arxiv]{1803.05457}~[cs.AI]
\urldef\tempurl%
\url{https://arxiv.org/abs/1803.05457}
\showURL{%
\tempurl}


\bibitem[Cobbe et~al\mbox{.}(2021)]%
        {gsm8k}
\bibfield{author}{\bibinfo{person}{Karl Cobbe}, \bibinfo{person}{Vineet Kosaraju}, \bibinfo{person}{Mohammad Bavarian}, \bibinfo{person}{Mark Chen}, \bibinfo{person}{Heewoo Jun}, \bibinfo{person}{Lukasz Kaiser}, \bibinfo{person}{Matthias Plappert}, \bibinfo{person}{Jerry Tworek}, \bibinfo{person}{Jacob Hilton}, \bibinfo{person}{Reiichiro Nakano}, \bibinfo{person}{Christopher Hesse}, {and} \bibinfo{person}{John Schulman}.} \bibinfo{year}{2021}\natexlab{}.
\newblock \showarticletitle{Training Verifiers to Solve Math Word Problems}.
\newblock  (\bibinfo{year}{2021}).
\newblock
\showeprint[arxiv]{2110.14168}~[cs.LG]
\urldef\tempurl%
\url{https://arxiv.org/abs/2110.14168}
\showURL{%
\tempurl}


\bibitem[Dua et~al\mbox{.}(2019)]%
        {drop}
\bibfield{author}{\bibinfo{person}{Dheeru Dua}, \bibinfo{person}{Yizhong Wang}, \bibinfo{person}{Pradeep Dasigi}, \bibinfo{person}{Gabriel Stanovsky}, \bibinfo{person}{Sameer Singh}, {and} \bibinfo{person}{Matt Gardner}.} \bibinfo{year}{2019}\natexlab{}.
\newblock \showarticletitle{{DROP}: A Reading Comprehension Benchmark Requiring Discrete Reasoning Over Paragraphs}. In \bibinfo{booktitle}{\emph{Proceedings of the 2019 Conference of the North {A}merican Chapter of the Association for Computational Linguistics: Human Language Technologies, Volume 1 (Long and Short Papers)}}, \bibfield{editor}{\bibinfo{person}{Jill Burstein}, \bibinfo{person}{Christy Doran}, {and} \bibinfo{person}{Thamar Solorio}} (Eds.). \bibinfo{publisher}{Association for Computational Linguistics}, \bibinfo{address}{Minneapolis, Minnesota}, \bibinfo{pages}{2368--2378}.
\newblock
\href{https://doi.org/10.18653/v1/N19-1246}{doi:\nolinkurl{10.18653/v1/N19-1246}}


\bibitem[Gaber et~al\mbox{.}(2024)]%
        {medrag}
\bibfield{author}{\bibinfo{person}{Farieda Gaber}, \bibinfo{person}{Maqsood Shaik}, \bibinfo{person}{Vedran Franke}, {and} \bibinfo{person}{Altuna Akalin}.} \bibinfo{year}{2024}\natexlab{}.
\newblock \showarticletitle{Evaluating large language model workflows in clinical decision support: referral, triage, and diagnosis}.
\newblock \bibinfo{journal}{\emph{medRxiv}} (\bibinfo{year}{2024}).
\newblock
\href{https://doi.org/10.1101/2024.09.27.24314505}{doi:\nolinkurl{10.1101/2024.09.27.24314505}}


\bibitem[Garza et~al\mbox{.}(2019)]%
        {blbp}
\bibfield{author}{\bibinfo{person}{Elba Garza}, \bibinfo{person}{Samira Mirbagher-Ajorpaz}, \bibinfo{person}{Tahsin~Ahmad Khan}, {and} \bibinfo{person}{Daniel~A. Jim\'{e}nez}.} \bibinfo{year}{2019}\natexlab{}.
\newblock \showarticletitle{Bit-level perceptron prediction for indirect branches}. In \bibinfo{booktitle}{\emph{Proceedings of the 46th International Symposium on Computer Architecture}} (Phoenix, Arizona) \emph{(\bibinfo{series}{ISCA '19})}. \bibinfo{publisher}{Association for Computing Machinery}, \bibinfo{address}{New York, NY, USA}, \bibinfo{pages}{27–38}.
\newblock
\showISBNx{9781450366694}
\href{https://doi.org/10.1145/3307650.3322217}{doi:\nolinkurl{10.1145/3307650.3322217}}


\bibitem[Gekhman et~al\mbox{.}(2024)]%
        {gekhman2024-fine}
\bibfield{author}{\bibinfo{person}{Zorik Gekhman}, \bibinfo{person}{Gal Yona}, \bibinfo{person}{Roee Aharoni}, \bibinfo{person}{Matan Eyal}, \bibinfo{person}{Amir Feder}, \bibinfo{person}{Roi Reichart}, {and} \bibinfo{person}{Jonathan Herzig}.} \bibinfo{year}{2024}\natexlab{}.
\newblock \showarticletitle{Does Fine-Tuning {LLM}s on New Knowledge Encourage Hallucinations?}. In \bibinfo{booktitle}{\emph{Proceedings of the 2024 Conference on Empirical Methods in Natural Language Processing}}, \bibfield{editor}{\bibinfo{person}{Yaser Al-Onaizan}, \bibinfo{person}{Mohit Bansal}, {and} \bibinfo{person}{Yun-Nung Chen}} (Eds.). \bibinfo{publisher}{Association for Computational Linguistics}, \bibinfo{address}{Miami, Florida, USA}, \bibinfo{pages}{7765--7784}.
\newblock
\href{https://doi.org/10.18653/v1/2024.emnlp-main.444}{doi:\nolinkurl{10.18653/v1/2024.emnlp-main.444}}


\bibitem[Gober et~al\mbox{.}(2022)]%
        {champsim}
\bibfield{author}{\bibinfo{person}{Nathan Gober}, \bibinfo{person}{Gino Chacon}, \bibinfo{person}{Lei Wang}, \bibinfo{person}{Paul~V. Gratz}, \bibinfo{person}{Daniel~A. Jimenez}, \bibinfo{person}{Elvira Teran}, \bibinfo{person}{Seth Pugsley}, {and} \bibinfo{person}{Jinchun Kim}.} \bibinfo{year}{2022}\natexlab{}.
\newblock \bibinfo{title}{The Championship Simulator: Architectural Simulation for Education and Competition}.
\newblock
\showeprint[arxiv]{2210.14324}~[cs.AR]
\urldef\tempurl%
\url{https://arxiv.org/abs/2210.14324}
\showURL{%
\tempurl}


\bibitem[{Google Research}(2025)]%
        {google_research_cache_replacement}
\bibfield{author}{\bibinfo{person}{{Google Research}}.} \bibinfo{year}{2025}\natexlab{}.
\newblock \bibinfo{title}{google-research: cache\_replacement}.
\newblock
\urldef\tempurl%
\url{https://github.com/google-research/google-research/tree/master/cache_replacement}
\showURL{%
\tempurl}
\newblock
\shownote{GitHub repository}.


\bibitem[Hegselmann et~al\mbox{.}(2025)]%
        {ehr_ralm}
\bibfield{author}{\bibinfo{person}{Stefan Hegselmann}, \bibinfo{person}{Georg von Arnim}, \bibinfo{person}{Tillmann Rheude}, \bibinfo{person}{Noel Kronenberg}, \bibinfo{person}{David Sontag}, \bibinfo{person}{Gerhard Hindricks}, \bibinfo{person}{Roland Eils}, {and} \bibinfo{person}{Benjamin Wild}.} \bibinfo{year}{2025}\natexlab{}.
\newblock \bibinfo{title}{Large Language Models are Powerful Electronic Health Record Encoders}.
\newblock
\showeprint[arxiv]{2502.17403}~[cs.LG]
\urldef\tempurl%
\url{https://arxiv.org/abs/2502.17403}
\showURL{%
\tempurl}


\bibitem[Hendrycks et~al\mbox{.}(2021)]%
        {math}
\bibfield{author}{\bibinfo{person}{Dan Hendrycks}, \bibinfo{person}{Collin Burns}, \bibinfo{person}{Saurav Kadavath}, \bibinfo{person}{Akul Arora}, \bibinfo{person}{Steven Basart}, \bibinfo{person}{Eric Tang}, \bibinfo{person}{Dawn Song}, {and} \bibinfo{person}{Jacob Steinhardt}.} \bibinfo{year}{2021}\natexlab{}.
\newblock \showarticletitle{Measuring Mathematical Problem Solving With the MATH Dataset}.
\newblock  (\bibinfo{year}{2021}).
\newblock
\showeprint[arxiv]{2103.03874}~[cs.LG]
\urldef\tempurl%
\url{https://arxiv.org/abs/2103.03874}
\showURL{%
\tempurl}


\bibitem[Hu et~al\mbox{.}(2021)]%
        {lora}
\bibfield{author}{\bibinfo{person}{Edward~J. Hu}, \bibinfo{person}{Yelong Shen}, \bibinfo{person}{Phillip Wallis}, \bibinfo{person}{Zeyuan Allen-Zhu}, \bibinfo{person}{Yuanzhi Li}, \bibinfo{person}{Shean Wang}, \bibinfo{person}{Lu Wang}, {and} \bibinfo{person}{Weizhu Chen}.} \bibinfo{year}{2021}\natexlab{}.
\newblock \bibinfo{title}{LoRA: Low-Rank Adaptation of Large Language Models}.
\newblock
\showeprint[arxiv]{2106.09685}~[cs.CL]
\urldef\tempurl%
\url{https://arxiv.org/abs/2106.09685}
\showURL{%
\tempurl}


\bibitem[Jain and Lin(2016)]%
        {hawkeye}
\bibfield{author}{\bibinfo{person}{Akanksha Jain} {and} \bibinfo{person}{Calvin Lin}.} \bibinfo{year}{2016}\natexlab{}.
\newblock \showarticletitle{Back to the Future: Leveraging Belady's Algorithm for Improved Cache Replacement}. In \bibinfo{booktitle}{\emph{Proceedings of the 43rd International Symposium on Computer Architecture (ISCA)}}. \bibinfo{publisher}{IEEE Press}, \bibinfo{address}{Seoul, Republic of Korea}, \bibinfo{pages}{78--89}.
\newblock
\href{https://doi.org/10.1109/ISCA.2016.17}{doi:\nolinkurl{10.1109/ISCA.2016.17}}


\bibitem[Jaleel et~al\mbox{.}(2010)]%
        {rrip}
\bibfield{author}{\bibinfo{person}{Aamer Jaleel}, \bibinfo{person}{Kevin~B. Theobald}, \bibinfo{person}{Simon~C. Steely}, {and} \bibinfo{person}{Joel Emer}.} \bibinfo{year}{2010}\natexlab{}.
\newblock \showarticletitle{High performance cache replacement using re-reference interval prediction (RRIP)}.
\newblock \bibinfo{journal}{\emph{SIGARCH Comput. Archit. News}} \bibinfo{volume}{38}, \bibinfo{number}{3} (\bibinfo{date}{June} \bibinfo{year}{2010}), \bibinfo{pages}{60–71}.
\newblock
\showISSN{0163-5964}
\href{https://doi.org/10.1145/1816038.1815971}{doi:\nolinkurl{10.1145/1816038.1815971}}


\bibitem[Jim\'{e}nez and Teran(2017)]%
        {mpppb}
\bibfield{author}{\bibinfo{person}{Daniel~A. Jim\'{e}nez} {and} \bibinfo{person}{Elvira Teran}.} \bibinfo{year}{2017}\natexlab{}.
\newblock \showarticletitle{Multiperspective reuse prediction}. In \bibinfo{booktitle}{\emph{Proceedings of the 50th Annual IEEE/ACM International Symposium on Microarchitecture}} (Cambridge, Massachusetts) \emph{(\bibinfo{series}{MICRO-50 '17})}. \bibinfo{publisher}{Association for Computing Machinery}, \bibinfo{address}{New York, NY, USA}, \bibinfo{pages}{436–448}.
\newblock
\showISBNx{9781450349529}
\href{https://doi.org/10.1145/3123939.3123942}{doi:\nolinkurl{10.1145/3123939.3123942}}


\bibitem[Jiménez and Lin(2001)]%
        {perceptron}
\bibfield{author}{\bibinfo{person}{Daniel~A. Jiménez} {and} \bibinfo{person}{Calvin Lin}.} \bibinfo{year}{2001}\natexlab{}.
\newblock \showarticletitle{Dynamic Branch Prediction with Perceptrons}. In \bibinfo{booktitle}{\emph{Proceedings of the Seventh International Symposium on High-Performance Computer Architecture (HPCA)}}. \bibinfo{publisher}{IEEE}, \bibinfo{pages}{197--206}.
\newblock
\href{https://doi.org/10.1109/HPCA.2001.903263}{doi:\nolinkurl{10.1109/HPCA.2001.903263}}


\bibitem[Karpukhin et~al\mbox{.}(2020)]%
        {dpr}
\bibfield{author}{\bibinfo{person}{Vladimir Karpukhin}, \bibinfo{person}{Barlas Oğuz}, \bibinfo{person}{Sewon Min}, \bibinfo{person}{Patrick Lewis}, \bibinfo{person}{Ledell Wu}, \bibinfo{person}{Sergey Edunov}, \bibinfo{person}{Danqi Chen}, {and} \bibinfo{person}{Wen tau Yih}.} \bibinfo{year}{2020}\natexlab{}.
\newblock \bibinfo{title}{Dense Passage Retrieval for Open-Domain Question Answering}.
\newblock
\showeprint[arxiv]{2004.04906}~[cs.CL]
\urldef\tempurl%
\url{https://arxiv.org/abs/2004.04906}
\showURL{%
\tempurl}


\bibitem[Khattab et~al\mbox{.}(2023)]%
        {dspy}
\bibfield{author}{\bibinfo{person}{Omar Khattab}, \bibinfo{person}{Arnav Singhvi}, \bibinfo{person}{Paridhi Maheshwari}, \bibinfo{person}{Zhiyuan Zhang}, \bibinfo{person}{Keshav Santhanam}, \bibinfo{person}{Sri Vardhamanan}, \bibinfo{person}{Saiful Haq}, \bibinfo{person}{Ashutosh Sharma}, \bibinfo{person}{Thomas~T. Joshi}, \bibinfo{person}{Hanna Moazam}, \bibinfo{person}{Heather Miller}, \bibinfo{person}{Matei Zaharia}, {and} \bibinfo{person}{Christopher Potts}.} \bibinfo{year}{2023}\natexlab{}.
\newblock \bibinfo{title}{DSPy: Compiling Declarative Language Model Calls into Self-Improving Pipelines}.
\newblock
\showeprint[arxiv]{2310.03714}~[cs.CL]
\urldef\tempurl%
\url{https://arxiv.org/abs/2310.03714}
\showURL{%
\tempurl}


\bibitem[Kim and Song(2024)]%
        {kim2024genie}
\bibfield{author}{\bibinfo{person}{Youngin Kim} {and} \bibinfo{person}{William~J. Song}.} \bibinfo{year}{2024}\natexlab{}.
\newblock \showarticletitle{Genie Cache: Non-Blocking Miss Handling and Replacement in Page-Table-Based DRAM Cache}. In \bibinfo{booktitle}{\emph{Proceedings of the 57th IEEE/ACM International Symposium on Microarchitecture (MICRO)}}. \bibinfo{publisher}{IEEE/ACM}, \bibinfo{pages}{983--996}.
\newblock
\href{https://doi.org/10.1109/MICRO61859.2024.00076}{doi:\nolinkurl{10.1109/MICRO61859.2024.00076}}


\bibitem[Liu et~al\mbox{.}(2020)]%
        {parrot}
\bibfield{author}{\bibinfo{person}{Evan~Zheran Liu}, \bibinfo{person}{Milad Hashemi}, \bibinfo{person}{Kevin Swersky}, \bibinfo{person}{Parthasarathy Ranganathan}, {and} \bibinfo{person}{Junwhan Ahn}.} \bibinfo{year}{2020}\natexlab{}.
\newblock \showarticletitle{An imitation learning approach for cache replacement}. In \bibinfo{booktitle}{\emph{Proceedings of the 37th International Conference on Machine Learning}} \emph{(\bibinfo{series}{ICML'20})}. \bibinfo{publisher}{JMLR.org}, Article \bibinfo{articleno}{579}, \bibinfo{numpages}{11}~pages.
\newblock
\href{https://doi.org/10.1145/3352460.3358319}{doi:\nolinkurl{10.1145/3352460.3358319}}


\bibitem[Liu(2022)]%
        {llamaindex}
\bibfield{author}{\bibinfo{person}{Jerry Liu}.} \bibinfo{year}{2022}\natexlab{}.
\newblock \bibinfo{booktitle}{\emph{{LlamaIndex}}}.
\newblock
\urldef\tempurl%
\url{https://github.com/jerryjliu/llama_index}
\showURL{%
\tempurl}


\bibitem[Min et~al\mbox{.}(2025)]%
        {graphrag}
\bibfield{author}{\bibinfo{person}{Congmin Min}, \bibinfo{person}{Rhea Mathew}, \bibinfo{person}{Joyce Pan}, \bibinfo{person}{Sahil Bansal}, \bibinfo{person}{Abbas Keshavarzi}, {and} \bibinfo{person}{Amar~Viswanathan Kannan}.} \bibinfo{year}{2025}\natexlab{}.
\newblock \bibinfo{title}{Efficient Knowledge Graph Construction and Retrieval from Unstructured Text for Large-Scale RAG Systems}.
\newblock
\showeprint[arxiv]{2507.03226}~[cs.AI]
\urldef\tempurl%
\url{https://arxiv.org/abs/2507.03226}
\showURL{%
\tempurl}


\bibitem[Mostofi et~al\mbox{.}(2025)]%
        {pacipv}
\bibfield{author}{\bibinfo{person}{Saba Mostofi}, \bibinfo{person}{Setu Gupta}, \bibinfo{person}{Ahmad Hassani}, \bibinfo{person}{Krishnam Tibrewala}, \bibinfo{person}{Elvira Teran}, \bibinfo{person}{Paul~V. Gratz}, {and} \bibinfo{person}{Daniel~A. Jim\'{e}nez}.} \bibinfo{year}{2025}\natexlab{}.
\newblock \showarticletitle{Light-weight Cache Replacement for Instruction Heavy Workloads}. In \bibinfo{booktitle}{\emph{Proceedings of the 52nd Annual International Symposium on Computer Architecture}} \emph{(\bibinfo{series}{ISCA '25})}. \bibinfo{publisher}{Association for Computing Machinery}, \bibinfo{address}{New York, NY, USA}, \bibinfo{pages}{1005–1019}.
\newblock
\showISBNx{9798400712616}
\href{https://doi.org/10.1145/3695053.3730993}{doi:\nolinkurl{10.1145/3695053.3730993}}


\bibitem[Packer et~al\mbox{.}(2024)]%
        {memgpt}
\bibfield{author}{\bibinfo{person}{Charles Packer}, \bibinfo{person}{Sarah Wooders}, \bibinfo{person}{Kevin Lin}, \bibinfo{person}{Vivian Fang}, \bibinfo{person}{Shishir~G. Patil}, \bibinfo{person}{Ion Stoica}, {and} \bibinfo{person}{Joseph~E. Gonzalez}.} \bibinfo{year}{2024}\natexlab{}.
\newblock \bibinfo{title}{MemGPT: Towards LLMs as Operating Systems}.
\newblock
\showeprint[arxiv]{2310.08560}~[cs.AI]
\urldef\tempurl%
\url{https://arxiv.org/abs/2310.08560}
\showURL{%
\tempurl}


\bibitem[Prakash et~al\mbox{.}(2025)]%
        {quarch}
\bibfield{author}{\bibinfo{person}{Shvetank Prakash}, \bibinfo{person}{Andrew Cheng}, \bibinfo{person}{Jason Yik}, \bibinfo{person}{Arya Tschand}, \bibinfo{person}{Radhika Ghosal}, \bibinfo{person}{Ikechukwu Uchendu}, \bibinfo{person}{Jessica Quaye}, \bibinfo{person}{Jeffrey Ma}, \bibinfo{person}{Shreyas Grampurohit}, \bibinfo{person}{Sofia Giannuzzi}, \bibinfo{person}{Arnav Balyan}, \bibinfo{person}{Fin Amin}, \bibinfo{person}{Aadya Pipersenia}, \bibinfo{person}{Yash Choudhary}, \bibinfo{person}{Ankita Nayak}, \bibinfo{person}{Amir Yazdanbakhsh}, {and} \bibinfo{person}{Vijay~Janapa Reddi}.} \bibinfo{year}{2025}\natexlab{}.
\newblock \bibinfo{title}{QuArch: A Question-Answering Dataset for AI Agents in Computer Architecture}.
\newblock
\showeprint[arxiv]{2501.01892}~[cs.AR]
\urldef\tempurl%
\url{https://arxiv.org/abs/2501.01892}
\showURL{%
\tempurl}


\bibitem[Qureshi et~al\mbox{.}(2007)]%
        {dip}
\bibfield{author}{\bibinfo{person}{Moinuddin~K. Qureshi}, \bibinfo{person}{Aamer Jaleel}, \bibinfo{person}{Yale~N. Patt}, \bibinfo{person}{Simon~C. Steely}, {and} \bibinfo{person}{Joel Emer}.} \bibinfo{year}{2007}\natexlab{}.
\newblock \showarticletitle{Adaptive insertion policies for high performance caching}.
\newblock \bibinfo{journal}{\emph{SIGARCH Comput. Archit. News}} \bibinfo{volume}{35}, \bibinfo{number}{2} (\bibinfo{date}{June} \bibinfo{year}{2007}), \bibinfo{pages}{381–391}.
\newblock
\showISSN{0163-5964}
\href{https://doi.org/10.1145/1273440.1250709}{doi:\nolinkurl{10.1145/1273440.1250709}}


\bibitem[Schick et~al\mbox{.}(2023)]%
        {toolformer}
\bibfield{author}{\bibinfo{person}{Timo Schick}, \bibinfo{person}{Jane Dwivedi-Yu}, \bibinfo{person}{Roberto Dess\'{\i}}, \bibinfo{person}{Roberta Raileanu}, \bibinfo{person}{Maria Lomeli}, \bibinfo{person}{Eric Hambro}, \bibinfo{person}{Luke Zettlemoyer}, \bibinfo{person}{Nicola Cancedda}, {and} \bibinfo{person}{Thomas Scialom}.} \bibinfo{year}{2023}\natexlab{}.
\newblock \bibinfo{title}{Toolformer: language models can teach themselves to use tools}.
\newblock \bibinfo{numpages}{13}~pages.
\newblock
\href{https://doi.org/10.5555/3666122.3669119}{doi:\nolinkurl{10.5555/3666122.3669119}}


\bibitem[Sethumurugan et~al\mbox{.}(2021)]%
        {rlr}
\bibfield{author}{\bibinfo{person}{Subhash Sethumurugan}, \bibinfo{person}{Jieming Yin}, {and} \bibinfo{person}{John Sartori}.} \bibinfo{year}{2021}\natexlab{}.
\newblock \showarticletitle{Designing a Cost-Effective Cache Replacement Policy using Machine Learning}. In \bibinfo{booktitle}{\emph{2021 IEEE International Symposium on High-Performance Computer Architecture (HPCA)}}. \bibinfo{pages}{291--303}.
\newblock
\href{https://doi.org/10.1109/HPCA51647.2021.00033}{doi:\nolinkurl{10.1109/HPCA51647.2021.00033}}


\bibitem[Shah et~al\mbox{.}(2022)]%
        {mockingjay}
\bibfield{author}{\bibinfo{person}{Ishan Shah}, \bibinfo{person}{Akanksha Jain}, {and} \bibinfo{person}{Calvin Lin}.} \bibinfo{year}{2022}\natexlab{}.
\newblock \showarticletitle{Effective Mimicry of Belady’s MIN Policy}. In \bibinfo{booktitle}{\emph{2022 IEEE International Symposium on High-Performance Computer Architecture (HPCA)}}. \bibinfo{pages}{558--572}.
\newblock
\href{https://doi.org/10.1109/HPCA53966.2022.00048}{doi:\nolinkurl{10.1109/HPCA53966.2022.00048}}


\bibitem[Shi et~al\mbox{.}(2019)]%
        {glider}
\bibfield{author}{\bibinfo{person}{Zhan Shi}, \bibinfo{person}{Xiangru Huang}, \bibinfo{person}{Akanksha Jain}, {and} \bibinfo{person}{Calvin Lin}.} \bibinfo{year}{2019}\natexlab{}.
\newblock \showarticletitle{Applying Deep Learning to the Cache Replacement Problem}. In \bibinfo{booktitle}{\emph{Proceedings of the 52nd Annual IEEE/ACM International Symposium on Microarchitecture}} (Columbus, OH, USA) \emph{(\bibinfo{series}{MICRO-52})}. \bibinfo{publisher}{Association for Computing Machinery}, \bibinfo{address}{New York, NY, USA}, \bibinfo{pages}{413–425}.
\newblock
\showISBNx{9781450369381}
\href{https://doi.org/10.1145/3352460.3358319}{doi:\nolinkurl{10.1145/3352460.3358319}}


\bibitem[Wei et~al\mbox{.}(2022)]%
        {emergence}
\bibfield{author}{\bibinfo{person}{Jason Wei}, \bibinfo{person}{Yi Tay}, \bibinfo{person}{Rishi Bommasani}, \bibinfo{person}{Colin Raffel}, \bibinfo{person}{Barret Zoph}, \bibinfo{person}{Sebastian Borgeaud}, \bibinfo{person}{Dani Yogatama}, \bibinfo{person}{Maarten Bosma}, \bibinfo{person}{Denny Zhou}, \bibinfo{person}{Donald Metzler}, \bibinfo{person}{Ed~H. Chi}, \bibinfo{person}{Tatsunori Hashimoto}, \bibinfo{person}{Oriol Vinyals}, \bibinfo{person}{Percy Liang}, \bibinfo{person}{Jeff Dean}, {and} \bibinfo{person}{William Fedus}.} \bibinfo{year}{2022}\natexlab{}.
\newblock \bibinfo{title}{Emergent Abilities of Large Language Models}.
\newblock
\showeprint[arxiv]{2206.07682}~[cs.CL]
\urldef\tempurl%
\url{https://arxiv.org/abs/2206.07682}
\showURL{%
\tempurl}


\bibitem[Wei et~al\mbox{.}(2023)]%
        {cot}
\bibfield{author}{\bibinfo{person}{Jason Wei}, \bibinfo{person}{Xuezhi Wang}, \bibinfo{person}{Dale Schuurmans}, \bibinfo{person}{Maarten Bosma}, \bibinfo{person}{Brian Ichter}, \bibinfo{person}{Fei Xia}, \bibinfo{person}{Ed Chi}, \bibinfo{person}{Quoc Le}, {and} \bibinfo{person}{Denny Zhou}.} \bibinfo{year}{2023}\natexlab{}.
\newblock \bibinfo{title}{Chain-of-Thought Prompting Elicits Reasoning in Large Language Models}.
\newblock
\showeprint[arxiv]{2201.11903}~[cs.CL]
\urldef\tempurl%
\url{https://arxiv.org/abs/2201.11903}
\showURL{%
\tempurl}


\bibitem[Wu et~al\mbox{.}(2011)]%
        {ship}
\bibfield{author}{\bibinfo{person}{Carole-Jean Wu}, \bibinfo{person}{Aamer Jaleel}, \bibinfo{person}{Will Hasenplaugh}, \bibinfo{person}{Margaret Martonosi}, \bibinfo{person}{Simon~C. Steely}, {and} \bibinfo{person}{Joel Emer}.} \bibinfo{year}{2011}\natexlab{}.
\newblock \showarticletitle{SHiP: signature-based hit predictor for high performance caching}. In \bibinfo{booktitle}{\emph{Proceedings of the 44th Annual IEEE/ACM International Symposium on Microarchitecture}} (Porto Alegre, Brazil) \emph{(\bibinfo{series}{MICRO-44})}. \bibinfo{publisher}{Association for Computing Machinery}, \bibinfo{address}{New York, NY, USA}, \bibinfo{pages}{430–441}.
\newblock
\showISBNx{9781450310536}
\href{https://doi.org/10.1145/2155620.2155671}{doi:\nolinkurl{10.1145/2155620.2155671}}


\bibitem[Yang et~al\mbox{.}(2025)]%
        {mlomics}
\bibfield{author}{\bibinfo{person}{Ziwei Yang}, \bibinfo{person}{Rikuto Kotoge}, \bibinfo{person}{Xihao Piao}, \bibinfo{person}{Zheng Chen}, \bibinfo{person}{Lingwei Zhu}, \bibinfo{person}{Peng Gao}, \bibinfo{person}{Yasuko Matsubara}, \bibinfo{person}{Yasushi Sakurai}, {and} \bibinfo{person}{Jimeng Sun}.} \bibinfo{year}{2025}\natexlab{}.
\newblock \bibinfo{title}{MLOmics: Cancer Multi-Omics Database for Machine Learning}.
\newblock
\showeprint[arxiv]{2409.02143}~[q-bio.GN]
\urldef\tempurl%
\url{https://arxiv.org/abs/2409.02143}
\showURL{%
\tempurl}


\bibitem[Yoo et~al\mbox{.}(2023)]%
        {stormbird}
\bibfield{author}{\bibinfo{person}{Ho~Jung Yoo}, \bibinfo{person}{Jeong~Hun Kim}, {and} \bibinfo{person}{Tae~Hee Han}.} \bibinfo{year}{2023}\natexlab{}.
\newblock \showarticletitle{RL-Based Cache Replacement: A Modern Interpretation of Belady’s Algorithm With Bypass Mechanism and Access Type Analysis}.
\newblock \bibinfo{journal}{\emph{IEEE Access}}  \bibinfo{volume}{11} (\bibinfo{year}{2023}), \bibinfo{pages}{145238--145253}.
\newblock
\href{https://doi.org/10.1109/ACCESS.2023.3346790}{doi:\nolinkurl{10.1109/ACCESS.2023.3346790}}


\end{thebibliography}

\appendix

\end{document}